\documentclass[a4paper,10pt]{article}
\usepackage[pdftex]{graphicx}
\PassOptionsToPackage{hyphens}{url}\usepackage{hyperref}
\hypersetup{
     colorlinks = true,
     allcolors = blue
}
\usepackage{amsmath,amsfonts,amssymb}
\usepackage{fullpage}
\usepackage{authblk}
\usepackage{setspace}
\usepackage{subcaption}
\usepackage{booktabs}
\usepackage{tabularx}
\usepackage{cancel}
\usepackage[normalem]{ulem}

\usepackage[explicit]{titlesec}
\usepackage{sidecap}
\usepackage{pbox}

\usepackage{placeins}
\usepackage{float}

\usepackage{amsfonts}
\usepackage{amsmath}
\usepackage{multirow}
\usepackage{longtable}
\usepackage{makecell}
\usepackage{graphicx}
\usepackage{xcolor}
\usepackage{comment}
\usepackage{hyperref}

\newlength{\figureDefaultWidth}
\newlength{\figureOneCol}
\newlength{\figureSmaller}
\date{}

\title{ \textbf{Heterogeneous rarity patterns drive price dynamics in NFT collections}}

\author[a]{Amin Mekacher}
\author[a]{Alberto Bracci}
\author[b]{Matthieu Nadini}
\author[c]{Mauro Martino}
\author[d]{Laura Alessandretti}
\author[e]{Luca Maria Aiello}
\author[a,f,g,*]{Andrea Baronchelli}

\affil[a]{\small Department of
Mathematics, City, University of London, London EC1V 0HB, UK}
\affil[b]{\small Elliptic Inc., London, UK}
\affil[c]{\small IBM Research, Cambridge MA, USA}
\affil[d]{\small Technical University of Denmark, DK-2800 Kgs. Lyngby, DK}
\affil[e]{IT University of Copenhagen, DK}
\affil[f]{\small UCL Centre for Blockchain Technologies, University
College London, London WC1E 6BT, UK}
\affil[g]{\small The Alan Turing Institute, London NW1 2DB, UK}

\affil[*]{\small Corresponding Author: abaronchelli@turing.ac.uk}

\begin{document}

\maketitle

\textbf{ 
We quantify Non Fungible Token (NFT) rarity and investigate how it impacts market behaviour by analysing a dataset of 3.7M transactions collected between January 2018 and June 2022, involving 1.4M NFTs distributed across 410 collections. First, we consider the rarity of an NFT based on the set of human-readable attributes it possesses and show that most collections present heterogeneous rarity patterns, with few rare NFTs and a large number of more common ones. Then, we analyze market performance and show that, on average, rarer NFTs: (i) sell for higher prices, (ii) are traded less frequently, (iii) guarantee higher returns on investment (ROIs), and (iv) are less risky, i.e., less prone to yield negative returns. 
We anticipate that these findings will be of interest to researchers as well as NFT creators, collectors, and traders.}

\section*{Introduction}

 Non Fungible Tokens, or NFTs, are digital titles (tokens) to property, either real or virtual, stored on a blockchain. They offer a powerful solution to long-standing issues related to the ownership of virtual and physical assets. They have swiftly revolutionised the art market \cite{art_nft}, the world of collectibles \cite{collectibles_nft}, the gaming industry \cite{gaming_nft}, and are promising to do the same with such sectors as luxury \cite{luxury_nft}, fashion \cite{fashion_nft}, music \cite{music_nft}, entertainment \cite{entertainment_nft}, and real-estate \cite{real_estate_nft,nadini2021mapping}. Throughout 2021, the NFT market grew by more than 61,000\%, starting from a monthly sale volume of 8,072,866 USD in January 2021 to 4,968,834,938 USD in January 2022~\cite{dune}. NFT was Collins Dictionary's word of the year for 2021~\cite{nft_collins}. 

NFT collections are groups of NFTs that share common features, such as visual aspects or the code that generated them \cite{generated_collections}. They have been a driving force for the booming NFT market~\cite{nadini2021mapping, nft_fortune}. In the prominent case of generative art, NFTs are associated to (virtual) objects made using a predetermined system, typically an algorithm, that often includes an element of chance~\cite{tate_generative}. To be concrete, CryptoPunks is a collection of 10,000 unique images of pixelated human faces algorithmically generated~\cite{cryptopunks}, while Bored Ape Yacht Club contains 10,000 profile pictures of cartoon apes that are generated by an algorithm~\cite{boredapesyacht}. Their market capitalization is 834M USD and 1.2B USD as of June 2022, respectively~\cite{raritytools}. 

NFTs in a collection are most often distinguishable from one another. For example, CryptoPunks have a gender (6,039 male and 3,840 female) and -- as for many other collections -- a number of \textit{traits} that distinguish them. So a punk can have, or not have, a ``Top Hat'', a ``Red Mohawk'', a ``Silver Chain'', or ``Wild White Hair'' among other possibilities. Furthermore, while most CryptoPunks are humans, there are also 88 Zombies, 24 Apes, and 9 Aliens in the collection. CryptoPunks are not equivalent according to the market. The most expensive CryptoPunk to date was sold for 23.7 million USD on February 12, 2022 \cite{highestpunk}, despite the average price of a punk being ``only'' 138,179 USD (see also~\cite{schaar2022non}). A similar picture holds for Bored Apes, with the most expensive one traded for 3.4 million USD on October 26, 2021~\cite{highestape}, vs an average price of 48,836 USD.

\begin{figure}[t]
    \centering
    \includegraphics[width=0.9\textwidth]{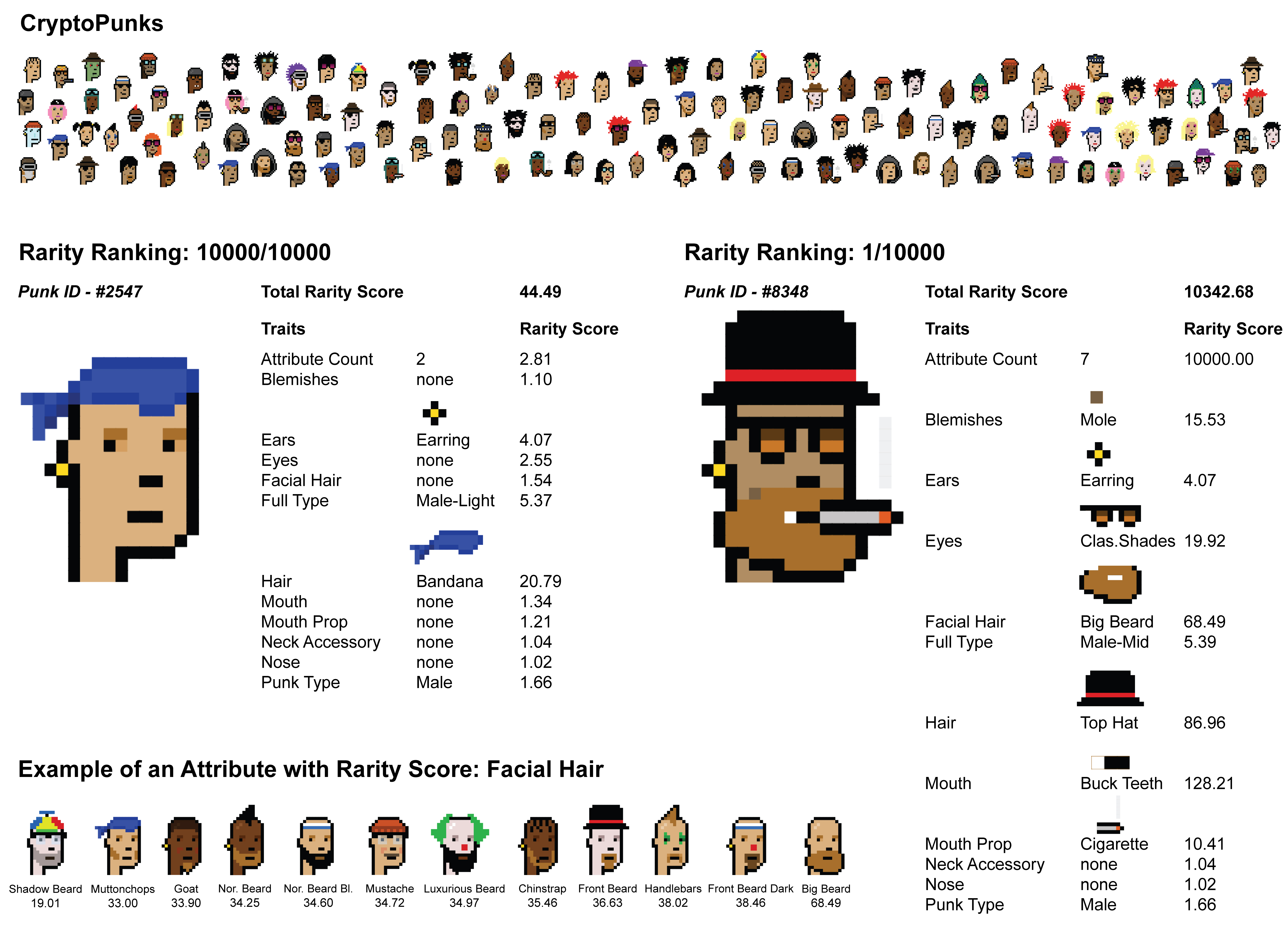}\\
    \caption{\textbf{Illustration of rarity in a collection.} 
    Demonstration of the visual difference between rare and common NFTs using the example of CryptoPunks. CryptoPunk \#2547 (on the right) is the least rare, as it has traits that appear frequently in the collection (i.e., the bandanna and the earring). CryptoPunk \#8348 is the rarest in the collection, mostly since it is the only one with seven non-null attributes. Rarity scores are not normalised. After normalisation, the total rarity score for punk \#2547 is zero, while the one for \#8348 is 100 (min and max of the collection, respectively). In the bottom left corner, we show, as an example, the rarity score of traits associated to the ``Facial Hair'' attribute. }
    \label{fig:cartoon}
\end{figure}

An hypothesis to rationalise these differences in price considers rarity. The heterogeneous distribution of traits among NFTs make some of them more rare than others (see Figure~\ref{fig:cartoon}), and scarcity is attractive for collectors \cite{koford1998market,lee2021measuring,hughes2020demand,ghazi2021preference}. However, despite some evidence that rarity and aesthetic preferences do play a role in the case of CryptoPunks \cite{kong2021alternative,schaar2022non}, a comprehensive analysis of the role of rarity on the market of NFTs is still lacking.

In this paper, we carry out a systematic investigation of how the rarity of NFTs impacts their market behaviour. 
We focus on within-collection rarity using the definition proposed in the platform rarity.tools \cite{raritytools}. Our dataset describes the rarity of 410 collections listed on OpenSea, containing a total of 1,479,020 NFTs that were exchanged 3,775,040 times between January 23, 2018 and June 6, 2022. First, we characterise trait distributions and investigate how they impact NFT rarity. Then, we analyse transaction data and find that: (i) rarity positively correlates with NFT prices and return on investments (ROIs), and (ii) negatively correlate with number of sales and risk quantified as the likelihood of a negative ROI. The breadth of our analysis suggests this market behaviour is likely to be genuinely self-organised. At the same time, our results could inform further research aimed at establishing how to optimally design collections, as well as effective trading strategies for the NFT market.

Our work adds to the relatively small yet rapidly growing body of research on the NFT ecosystem. 
Previous studies include an overview of the overall market, trade networks, and visual features of NFTs, and their impact on price prediction~\cite{nadini2021mapping}, as well as of the underlying technologies, such as the blockchain and the smart contracts, with a risk assessment~\cite{wang2021non}. Other research has focused generally on specific marketplaces or collections, analysing such issues as the determinants of success of NFT artists~\cite{barabasi2022foundation}, the role of social media attention~\cite{tweetprice}, the creators-collectors network~\cite{franceschet2021hits}, and the financial advantage of experienced users~\cite{investors2022}. Along this line, research also suggests that NFTs have become a promising investment as a fintech asset~\cite{bao2021recent}. Other lines of research include the analysis of illicit transactions connected to NFT trading~\cite{von2021nft,pelechrinis2022spotting} and of their connections with financial indicators~\cite{ante2021non,borri2022economics,aharon2021nfts}. The metaverse, an NFT submarket which has recently garnered attention both from big tech companies~\cite{zuckerbergmetaverse} and popular NFT creators~\cite{yugalabsmetaverse}, is another focus of research~\cite{ynag2022fusing,lee2021creators}. 

\section*{Background, data and methods}

\subsection*{Glossary of key terms}

\noindent \textit{NFTs.} An NFT -- or Non Fungible Token -- is a unit of data stored in a blockchain that certifies a digital asset to be unique and not interchangeable. It provides uncontroversial answers to such questions as who owns, previously owned, and created the NFT, as well as which of the many copies is the original. Several types of digital objects can be associated to an NFT including photos, videos, and audio. Crucially, NFTs can be sold and traded, and are used to commodify digital as well as non-digital objects in different contexts, such as art, gaming, music and fashion. Started on Ethereum \cite{ethhub}, today NFTs are available on several other blockchains. 

\noindent  \textit{Attributes and traits.} Attributes refer to human-readable characteristics of an NFT. In generative art, for example, they usually relate to visual properties of items. Attributes can take one among several values. For example, in the CryptoPunks collection, every item has the attribute ``type'' that can take one among the following traits: ``Male'', ``Female'', ``Zombie'', ``Ape'' or ``Alien''. CryptoPunks have also attributes that capture the presence of any accessory, such as earrings or bandanas. For the remainder of this study, we refer to the value taken by an attribute as the \textit{trait}.

\noindent \textit{Collections.} A collection is a group of NFTs whose associated digital items share common features. When minting an NFT, a creator can include the corresponding item within a collection. In generative art, for example, items of a collection are created by the same generative algorithm. 

\noindent \textit{Marketplaces.} Creators and collectors meet in online marketplaces to trade NFTs. The largest of these markets, OpenSea~\cite{opensea}, enables any creator to sell their NFTs and, at the moment of writing, it offers 44 million NFTs~\cite{opensea_assets}. Other marketplaces feature a curated selection of creators (e.g., Foundation~\cite{foundation_website}, SuperRare~\cite{superrare_website}, Nifty Gateway~\cite{nifty}, Feral File~\cite{feral}). NFTs are auctioned on these marketplaces, where the NFT can be sold to the highest bidder or with a declining price, depending on the kind of auction. 
After an NFT is minted on a marketplace -- i.e., it is converted into a digital asset on the blockchain -- it can be put up for auction. Typically,  the first transaction, from the creator of the NFT to the first user, is different in nature from the subsequent trades (e.g., the first user is often not able to select a specific NFT from a collection \cite{nft_mint_sale}). 

\subsection*{Dataset}
Our dataset includes 3,775,040 sales, taking place on the Ethereum blockchain, of 1,479,020 NFTs from 410 collections, including 61 of the top 100 collections by sales volume according to CoinMarketCap \cite{cmc}. The list of collections considered in this study can be found in the Supplementary Information. The dataset was built by considering all collections we could automatically match (by name) between rarity.tools \cite{raritytools} - a website dedicated to ranking collectible NFTs, also sometimes called Profile Picture NFT projects (PFP), by rarity - and the Opensea market. From the latter, we collected the release date, NFT traits and all sales concerning these collections that took place between January 23, 2018 and June 6, 2022. To avoid spurious effects, we only considered user-to-user transactions, where buyer and sellers are both aware of the precise identity of the traded NFT (i.e., we discarded the initial creator-to-user transactions). In the following, we refer to the first user-to-user transaction as ``primary'' sale, and to all subsequent transactions as ``secondary'' sales. Where not specified, by ``sales'' we consider both primary and secondary sales.

\begin{figure}[!t]
    \centering
    \includegraphics[width=0.9\textwidth]{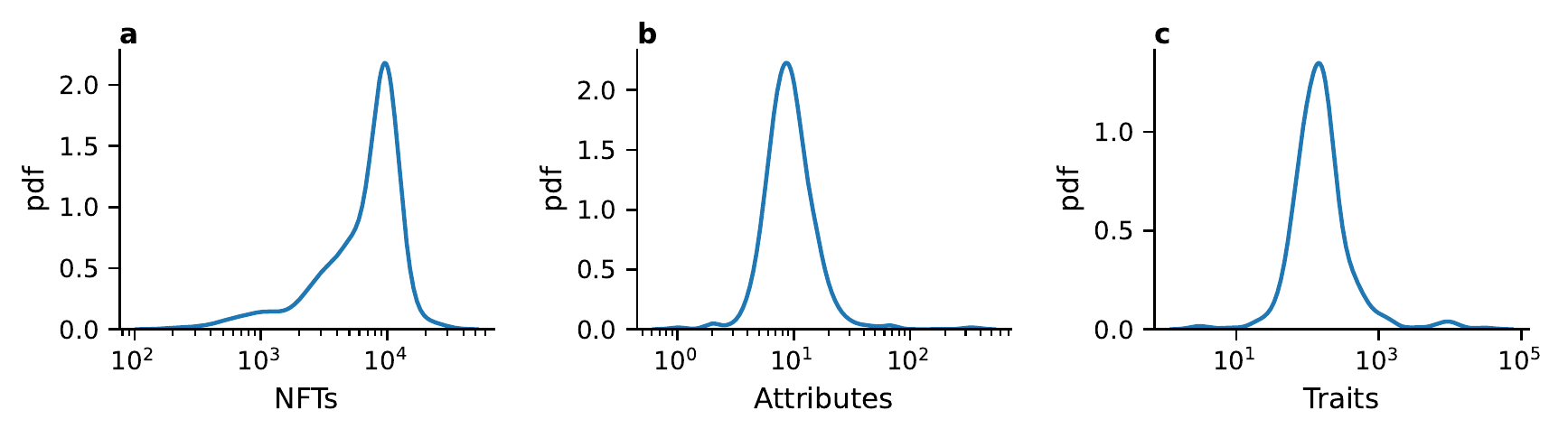}\\
    \caption{\textbf{Characteristics of collections.} The probability distribution of a) the total number of NFTs within the collection; b) the total number of attributes in the collection; c) the total number of traits in the collection.}
    \label{fig:collection_traits}
\end{figure}

Collections in our dataset have on average $7,554$ NFTs. There is, however, wide heterogeneity across collections: the standard deviation of the distribution is $194.64$, and the mode is $10,000$ NFTs (see Figure \ref{fig:collection_traits}a). The number of attributes in a collection is $11.1 \pm 0.91$, where the reported error corresponds to the standard deviation of the distribution (see Figure \ref{fig:collection_traits}b). As for the number of traits, the average is equal to $415.1$, with a standard deviation of $97.6$ (see Figure \ref{fig:collection_traits}c). On average, an attribute within a collection has $37.4$ different traits. More information about the algorithms used to assign traits to an NFT can be found in the Supplementary Information.

\subsection*{Rarity}

 The rarity of a trait is quantified as the fraction of NFTs within a collection having this trait. This value is indicated on OpenSea's sale page. 
For a collection containing $N$ NFTs, the \textit{trait rarity score}, $R_t$, for a trait $t$ shared by $r$ NFTs is defined as:

\begin{equation}
    R_t = \left({\frac{r}{N}}\right)^{-1}
\end{equation}

To quantify the overall rarity of an NFT within a collection, we consider each trait independently and define the \textit{NFT rarity score}, $R_{NFT}$, as the sum of the rarity scores of each one of its traits, that is 

\begin{equation}
R_{NFT} = \sum_t R_t.
\end{equation}

\noindent In order to compare this score between collections, we then normalize the scores within a collection with a min-max normalization. For a collection with a maximum and a minimum rarity score $R_{max}$ and $R_{min}$ respectively, the normalised rarity score $R_{norm}$ is given by $
    R_{norm} = 100 (R - R_{min}) / (R_{max} - R_{min}) $.
 By doing so, every NFT ends up with a normalised rarity score between $0$ (least rare) and $100$ (rarest). All the analyses presented in the main text of this article are based on the NFT rarity score.

Finally, we also consider \textit{the NFT rarity rank}, where the rarity rank of an NFT is given by the trait rarity rank of its rarest trait.
This metric will allow us to quantify the effect of a rare trait on the market behaviour of an NFT, regardless of its other traits. Analyses based on the NFT rarity rank can be found in the Supplementary Information.

Further information on measuring NFT rarity, including a detailed discussion of the above measures, can be found in~\cite{rarityscore}.

\section*{Results}

\subsection*{Market Growth}

\begin{figure}[t]
    \centering
    \includegraphics[width=0.8\textwidth]{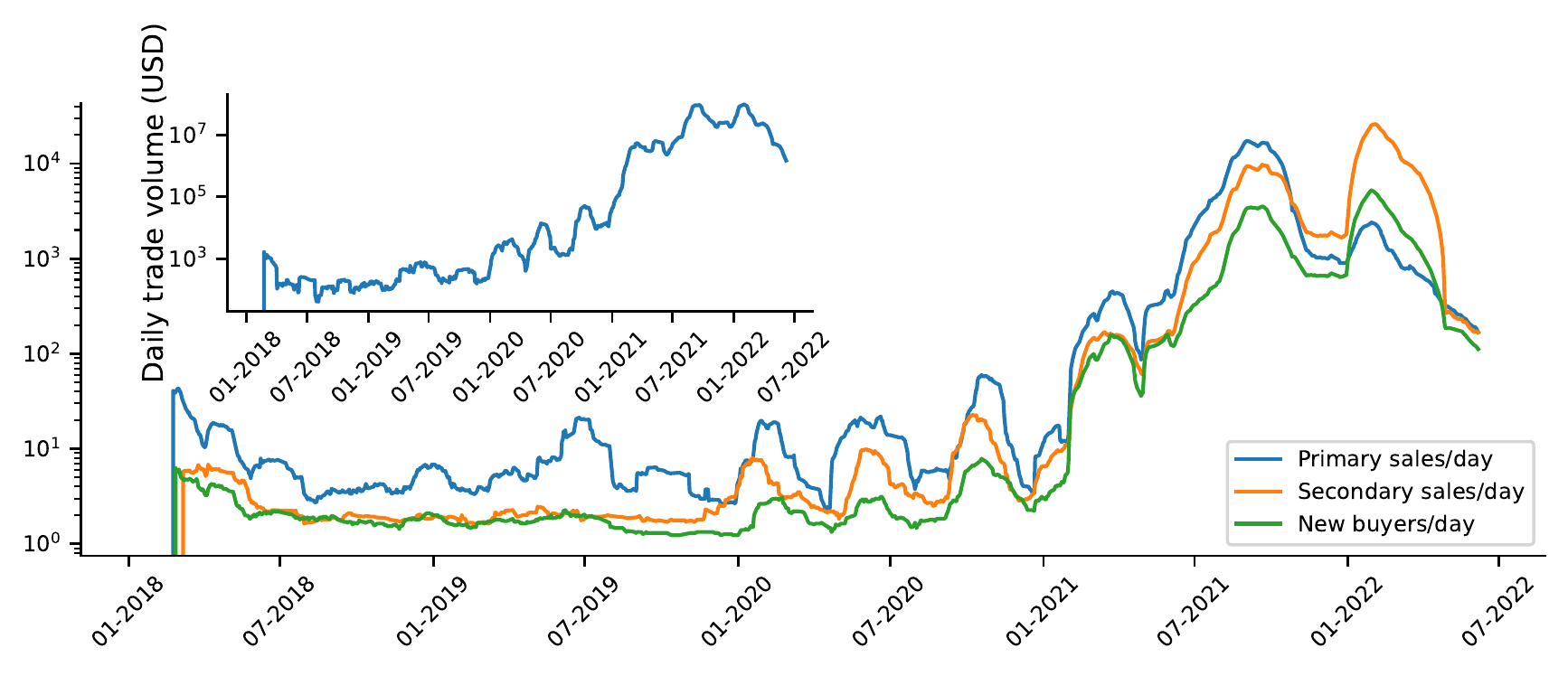}\\
    \caption{\textbf{The collectible market boom.}
    30-day rolling average of the daily number of primary sales (blue line), secondary sales (orange line) and new buyers (i.e., new wallet addresses buying an NFT for the first time, green line). Inset: 30-day rolling average of the daily trade volume (in USD).}
    \label{fig:crypto_boom}
\end{figure}

We start by investigating the evolution of NFT sales in our dataset over time. We find that the interest in the collections remained stable until the end of 2020, then started to gain traction in 2021, especially in terms of available NFTs on the market (see SI Fig.~1). The number of primary sales grew from an average of $14$ daily sales in January 2021 to $784$ sales every day in March 2022, when the market peaked, implying a percentage growth of $5,500\%$ (see Figure \ref{fig:crypto_boom}). Similarly, secondary sales grew by $110,177\%$, starting from $9$ sales/day in January 2021 and reaching $9925$ sales/day in March 2022. Interestingly, around October 2021, the number of secondary sales started to exceed the number of primary sales, a trend that still holds at the moment of writing. This surge in activity led to a growth of daily volume of trades of $18,520\%$ between January 2021 and March 2022 (see Figure \ref{fig:crypto_boom} inset), and attracted new users. The number of new buyers increased by $41,755\%$ in 2021. These results indicate an overall growth of the popularity of NFT collections on OpenSea, both with respect to the size of the NFT community, and to the total market value.

\begin{figure}[t]
    \centering
    \includegraphics[width=0.9\textwidth]{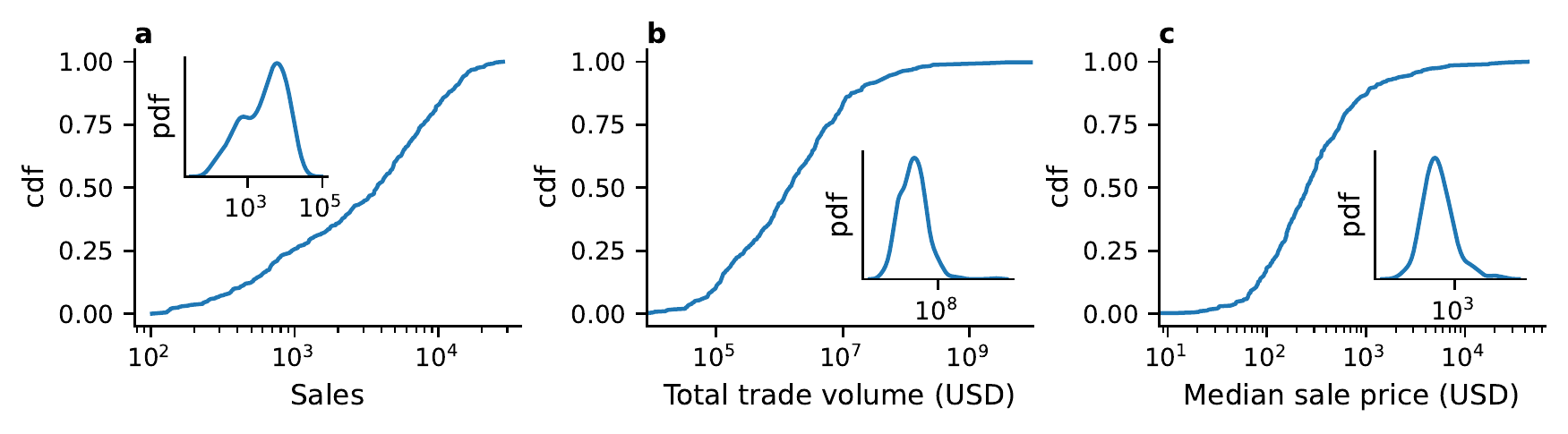}\\
    \caption{\textbf{Cumulative distribution of key market properties across collections.} a) Total number of sales per collection.  
    b) Total trade volume per collection. c) Median sale price per collection. In inset: Distribution of each market property.}
    \label{fig:collections_stats}
\end{figure}

Different collections contributed to varying extents to the growth of the collectible NFT market.
Figure~\ref{fig:collections_stats} shows the distribution of key market properties across NFT collections: total number of sales per collection (Figure~\ref{fig:collections_stats}a), total traded volume per collection (Figure~\ref{fig:collections_stats}b) and collection items median sale price (Figure~\ref{fig:collections_stats}c). 

Collections are widely heterogeneous with respect to market properties. $25.6\%$ of the collections have generated less than $1,000$ sales, whereas $17.1\%$ have generated more than $10,000$ (see Figure~\ref{fig:collections_stats}a).
Further, $43.9\%$ of the collections had a total trade volume below a million dollars, whereas $3.64\%$ generated more than a hundred million dollars of sales on the marketplace (see Figure~\ref{fig:collections_stats}b). The success of a collection can also be measured by looking at the median price at which its NFTs are sold on OpenSea. For $18.3\%$ of the collections, the median sale price is lower or equal to a hundred dollars, whereas it is higher than a thousand dollars for $12.9\%$ of the considered collections (see Figure~\ref{fig:collections_stats}c). These findings indicate that collectibles NFT do not meet the same success on OpenSea, a claim that is supported by the infamous success stories of a few collections, whereas the others quickly become a flop on the platform \cite{forbes_flop}.

\subsection*{Quantifying rarity} 

\begin{figure}[t]
    \centering
    \includegraphics[width=0.9\textwidth]{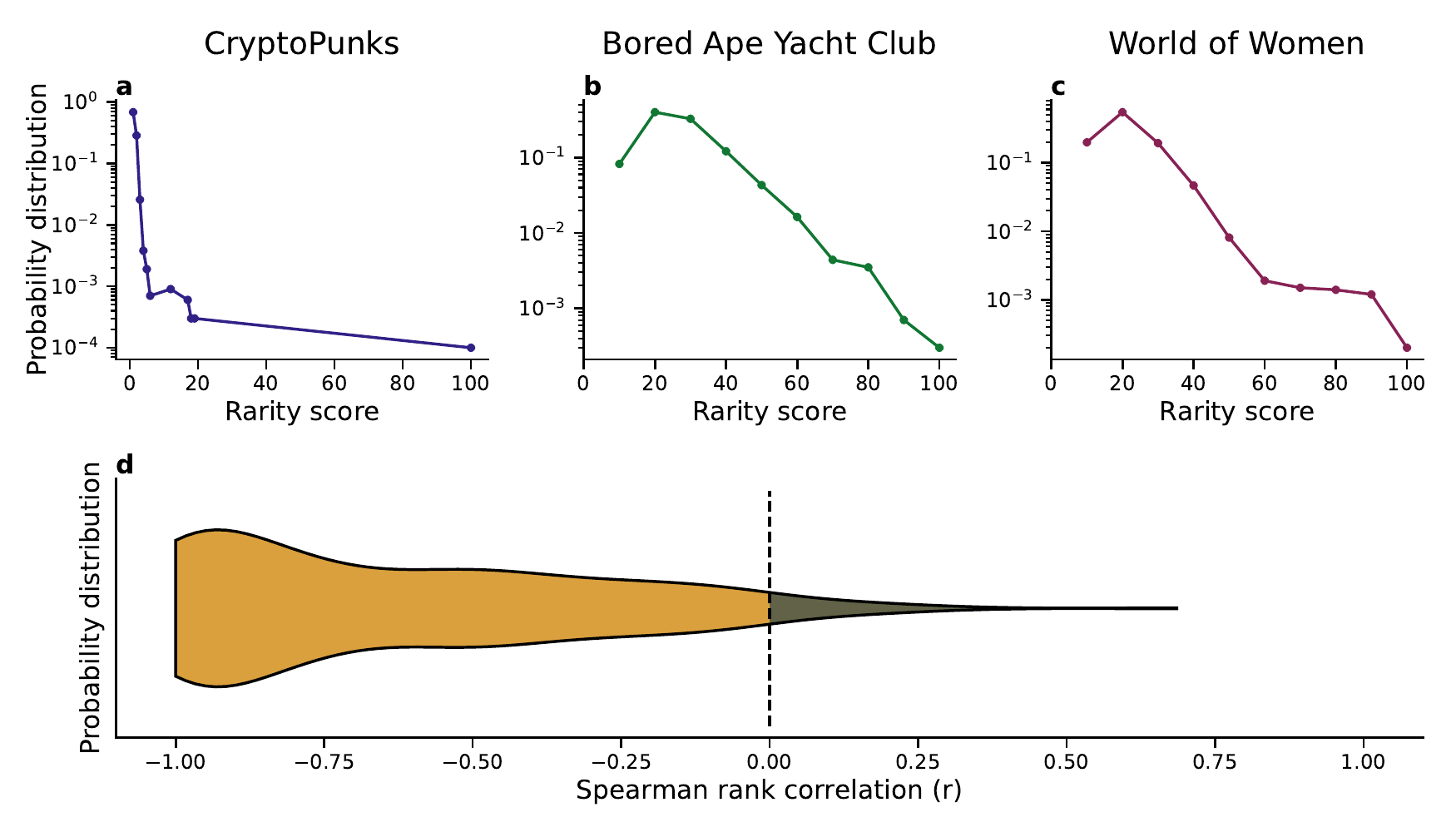}\\
    \caption{\textbf{Rare NFTs are scarce within a collection.} 
    a-c) Distribution of the rarity score of the NFTs within three collections: CryptoPunks (a), Bored Ape Yacht Club (b), and World of Women (c). d) Violin plot of the Spearman Rank correlation computed between the rarity score and the number of NFTs with that score. $96\%$ of the collections have a Spearman rank $r \leq 0$ (black dashed lines). }
    \label{fig:rarity_design_score}
\end{figure}

We quantify the distribution of rarity scores for items within the same collection. As an example, Figure~\ref{fig:rarity_design_score} shows the distribution of rarity for three popular collections, namely CryptoPunks, Bored Ape Yacht Club, and World of Women.

For CryptoPunks, the median rarity score is $0.82$, with only one of the $10,000$ CryptoPunks having a rarity score above $75$, whereas $99.7\%$ of the tokens have a rarity score below $10$ (see Figure~\ref{fig:rarity_design_score}a). Moreover, as most of the CryptoPunks have a low rarity score, the least rare ones are aggregated into two bins, whereas the rare one occupies the only bin with a high rarity score within the collection. The median rarity score for Bored Ape Yacht Club is $20.3$, and 26 apes (i.e., $0.26\%$ of the collection) have a rarity score above $75$. The distribution is skewed towards lower rarity scores, with $68.2\%$ of the assets with a rarity score below $25$, among which $8.23\%$ fall below a rarity score of $10$ (see Figure~\ref{fig:rarity_design_score}b). The profile for the World of Women collection is also not as heterogeneous as that of CryptoPunks; it has a median rarity score of $14.8$ and only 24 assets ($0.24\%$ of the collection) have a rarity score above $75$. $87.3\%$ of the tokens have a rarity score below $25$, and $19.9\%$ of those lie below a rarity score of $10$ (see Figure \ref{fig:rarity_design_score}c). To generalize these observations, we calculated the Spearman rank correlation coefficient between the rarity bin and the number of NFTs by rarity bin. A negative value of the correlation coefficient indicates that the higher the rarity score, the lower the supply of NFTs is within the considered collection. Like the three example collections in Figures~\ref{fig:rarity_design_score}a-c, $96\%$ of the collections in our dataset have a Spearman rank $r \leq 0$ , as shown in Figure~\ref{fig:rarity_design_score}d, where the violin plot represents the probability distribution of the Spearman rank correlation by collection. We compare the ability of $6$ different statistical distributions, namely the exponential, power-law, uniform, cauchy, log-normal and levy distributions, to capture the distribution of rarity for each collection, using the Akaike model selection method~\cite{wagenmakers2004aic} (see SI for more details). We find that, among the distributions considered, $90\%$ of the collections are best described by a log-normal distribution (with $\langle \mu \rangle = 0.91 \pm 0.16$, see SI Fig.~2), only $7\%$ by an exponential, $1\%$ by a uniform function and the rest by heterogeneous distributions such as power-laws or Levy (for a visualization of a sample of these distributions, see SI Fig.~3).

The same correlation analysis performed using the rarity rank confirms our results (see SI Fig.~4) 
In the following, we will focus on NFTs rarity score, because this metric takes into account all the traits associated with an NFT, and is therefore more suitable to quantify NFTs properties and rarity.
All the following results are replicated using trait rarity rank as robustness check (see Supplementary Information section E.1)

Our analysis indicates that the distribution of the rarity within a collection is heterogeneous, thus leading to a situation where rare NFTs are genuinely scarce on the marketplace. Notice that while this may seem trivial (``rare items are fewer than common items''), the distribution of traits rarity, and in turn their combination in single NFTs could in principle generate a wide range of distributions of NFT rarity, including homogeneous ones. 

\subsection*{Rarity and market performance}

To measure the relationship between rarity and market performance, we compute the rarity score of each NFT, and we split the assets into quantiles with respect to their rarity score to analyse collections individually. We then compare the median sale price across quantiles. We are using quantiles to ensure that NFTs within a collection will be evenly balanced between each bin, as to avoid having a collection skewing the results in the aggregated analysis, by having all of its NFTs concentrated in a single bin. For the individual collections analysis, NFTs are partitioned into twenty quantiles, whereas 100 quantiles are used when aggregating the collections together. 

\begin{figure}[t]
    \centering
    \includegraphics[width=0.9\textwidth]{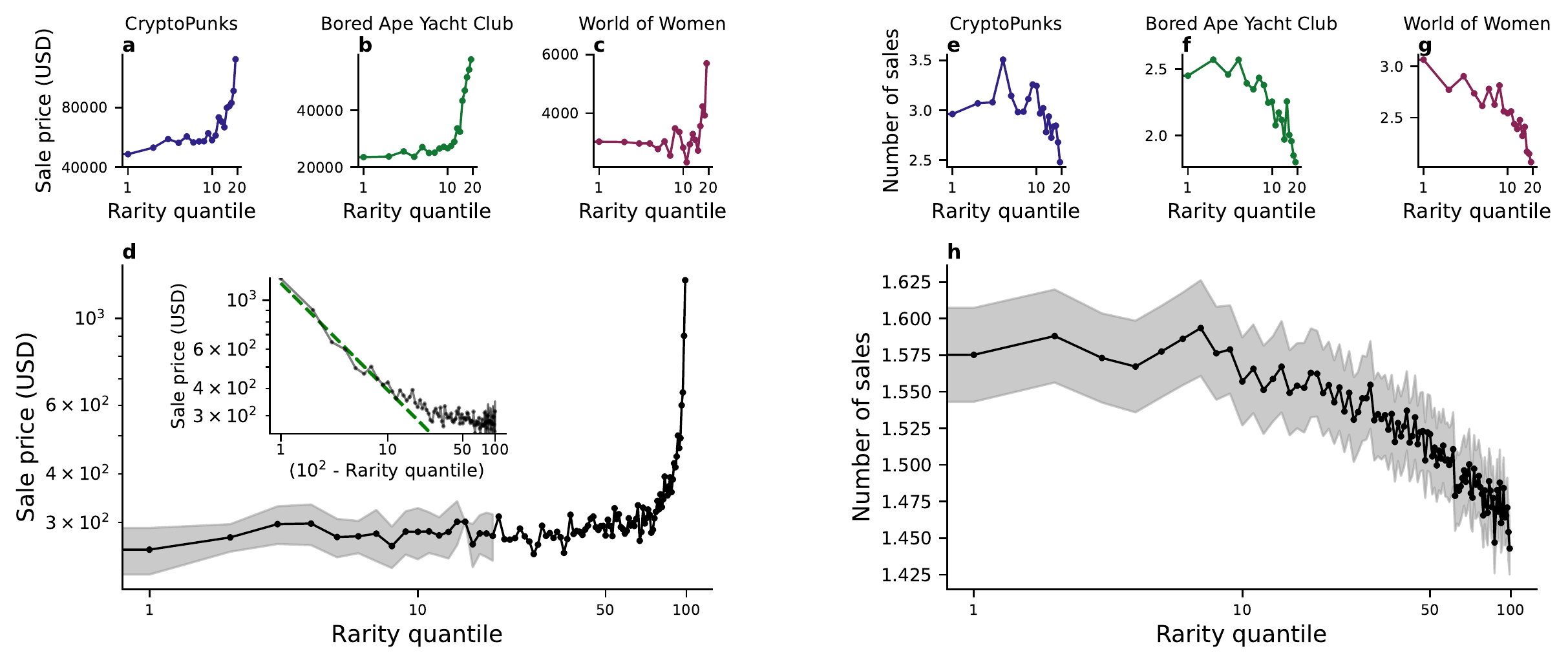}\\
    \caption{\textbf{Rare NFTs have a higher financial value and circulate less on the marketplace.} 
     Median sale price in USD (a-c) and average number of sales (e-g) by rarity quantile (with 20 quantiles considered) for three collections: CryptoPunks (a and e), Bored Ape Yacht Club (b and f), and World Women (c and g). d) Median sale price by rarity quantile (with 100 quantiles considered) considering all collections. Inset: median sale price against the quantity (100-$q$), where $q$ is the rarity quantile, in log-log scale (black line) and the corresponding power law fit (green dashed line). h) Median number of sales by rarity quantile considering all collections.}
    \label{fig:impact_rarity_agg}
\end{figure}

First, we consider the relation between market behaviour and rarity for three exemplar collections, CryptoPunks, Bored Ape Yacht
Club, and World of Women (see Figure \ref{fig:impact_rarity_agg}a-c). We observe that the median sale price at which NFTs are auctioned is relatively constant for the most common NFTs in each collection (rarity quantile smaller than 10), and then increase sharply for the rarest NFTs (rarity quantile larger than 10, see Figure \ref{fig:impact_rarity_agg}a-c). These findings are robust, and are observed also when we consider NFTs in all collections (see Figure \ref{fig:impact_rarity_agg}d). We notice that the median sale price is relatively flat for the $50\%$ least rare NFTs, before increasing by $195\%$ for the $10\%$ rarest NFTs. More strikingly, the median sale price for the $90\%$ least rare NFTs is equal to $298 \pm 3.2$ USD, and rises to $1,254$ USD for the $1\%$ rarest NFTs. Focusing on the top $10\%$ rarest NFTs, the relationship between the median sale price $p$ and the quantity (100 -$q$), where $q$ is the rarity quantile, is well described by a power law $p\sim (100 - q)^{\alpha}$ with exponent $\alpha = -0.55$ (see Figure \ref{fig:impact_rarity_agg} inset). This result indicates a strong relationship between NFT rarity and median sale price.

On the other side, we find that rare NFTs are not sold as frequently as common ones on the marketplaces. By looking at the individual collections, we see that the average number of sales decreases as we increase the rarity of the NFTs we are considering (see Figure \ref{fig:impact_rarity_agg}e-g). Regarding the average number of sales, by aggregating all collections together, we find that the number of sales decreases for rarer NFTs. In particular, the $1\%$ least rare NFTs are sold, on average, $10.8\%$ more than the $1\%$ rarest ones (see Figure \ref{fig:impact_rarity_agg}h).

In order to check that this behaviour holds when considering a shorter time span within OpenSea's lifetime, we performed the same analysis by considering only sales happening during the third quarter of 2021 (see SI Fig.~9) and the fourth quarter as well (see SI Fig.~11). Our findings are also robust by considering the sale price in ETH rather than in USD (see SI Fig.~7), and by discarding the rarest and least rare NFTs from each collection (see SI Fig.~13). Moreover, we notice a similar pattern when quantifying the rarity of the NFTs with the NFT rarity rank instead of the NFT rarity score (see SI Fig.~5).

\subsection*{Rarity and return on investment}

NFTs can be purchased and later put on sale again on the marketplace. An NFT owner is free to set an initial price to an auction, and to transfer the ownership of the NFT to the highest bidder. As such, NFTs which have been minted years ago, such as the CryptoPunks, can still be purchased on OpenSea. 
The results shown in Figure~\ref{fig:impact_rarity_agg} indicate that, within a collection, the rarest NFTs are typically sold at a higher absolute price than the least rare ones on the market. However, this fact does not necessarily imply that the return on investment of secondary sales is positive, as it does not take into account the price at which the asset was initially purchased before being auctioned again.
To study whether the correlation between rarity and price strengthens as a token keeps being exchanged on the market, we computed the return $R$ of the $k^{th}$ sale of an NFT as:

\begin{equation}
    R = \frac{P(k) - P(k-1)}{P(k-1)},
\end{equation}

\noindent where $P(k)$ is the price that was paid for the NFT for its $k^{th}$ sale. A positive return indicates that the NFT was sold at a higher price than the one it was bought for, whereas a negative return represents a financial loss for the seller. 

\begin{figure}[t]
    \centering
    \includegraphics[width=0.9\textwidth]{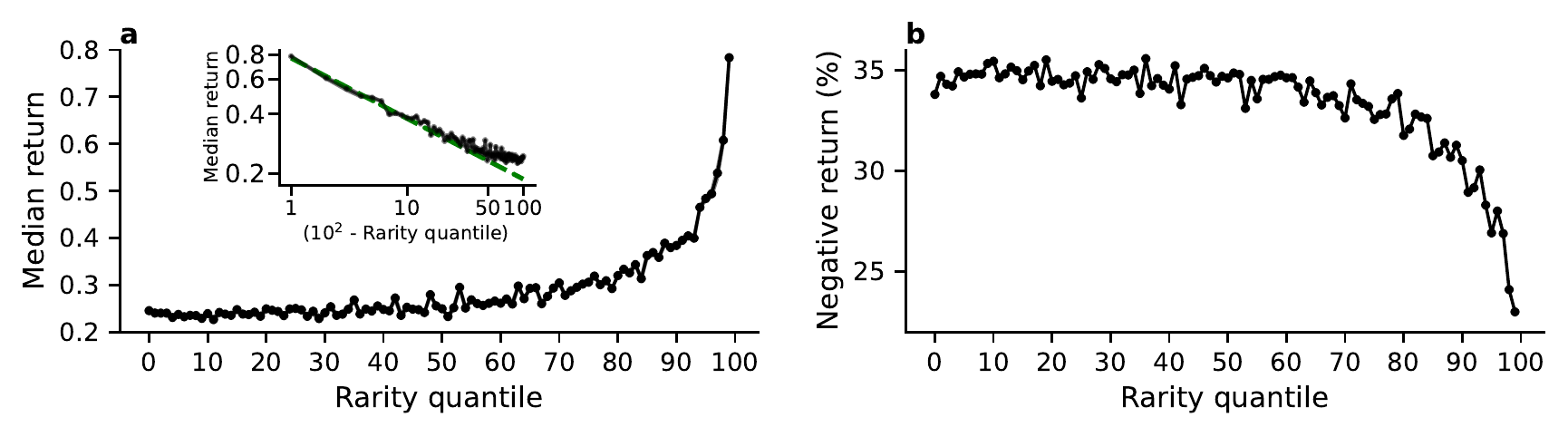}\\
    \caption{\textbf{High rarity leads to higher returns, and a lower chance of a negative return.} a) Median return in USD by rarity quantile. Inset: median return against the quantity (100-$q$), where $q$ is the rarity quantile in log-log scale (black line) and the corresponding power law fit (green dashed line). b) Fraction of sales with negative return in USD by rarity quantile.}
    \label{fig:amplification_neg_return}
\end{figure}

Figure~\ref{fig:amplification_neg_return}a shows the median return computed when aggregating all collections by rarity quantile. We find that the rarest NFTs have a much higher median return, whereas the value is almost constant in the first half of the curve. Focusing on the top $10\%$ rarest NFTs, we observe that the relationship between the quantity (100 - $q$), where $q$ is the rarity quantile, and the median return $R$ is well described by a power law $R\sim (100 - q)^{\alpha}$, with an exponent $\alpha = -0.29$ (see Figure \ref{fig:amplification_neg_return} inset). The median return is relatively flat around $0.24 \pm 0.001$ for the $50\%$ least rare NFTs, thus indicating no noticeable advantage for an NFT to be one of the least rare assets of the collection or an average one in terms of rarity, whereas the median return grows by $105\%$ within the top $10\%$ rarest NFTs. 
Finally, we study the relation between NFT rarity and the probability to generate negative returns.  We observe that, on average, rarer NFTs are less likely to generate negative returns (see Figure~\ref{fig:amplification_neg_return}b). The fraction of sales generating negative returns is equal to $34.6 \pm 0.58\%$ for the $50\%$ least rare NFTs, but drops from $30.5\%$ to $22.9\%$ within the top $10\%$ rarest NFTs, i.e., a decrease of $24.9\%$. These results also hold by only considering the sales happening during a shorter a shorter time period, such as the third quarter of 2021 (see SI Fig.~10) and the fourth quarter (see SI Fig.~12). The same analysis has been performed by considering the sale prices in ETH (see SI Fig.~8) and by discarding the rarest and least rare NFTs of every collection (see SI Fig.~14). These results are also robust when using the NFT rarity rank to measure the rarity of an NFT rather than the rarity score (see SI Fig.~6).

\subsection*{Discussion}
We have quantified rarity in 410 NFT collections and analysed its effect on market performance. Rarity is a fundamental feature of NFTs belonging to a collection because (i) it allows users to categorise NFTs on the traditionally market-relevant axis of scarcity and (ii) it is based on human-readable, easy to identify, traits that creators assign to NFTs. We have found that the distribution of rarity is heterogeneous throughout the vast majority of collections. We have shown that rarity is positively correlated with the sale price and negatively correlated with the number of sales of an NFT, with the effect being stronger for the top $10\%$ rare NFTs. Last, we have shown how rarity is associated with higher return of investment and lower probability of yielding negative returns in secondary sales.

The finding that most rarity distributions are heavily heterogeneous, with few very rare NFTs, is interesting since in principle more homogeneous distributions would be possible. The ubiquitous nature of this pattern may indicate either that creators deliberately choose heterogeneous distributions (design perspective) or that heterogeneous distributions help make a collection successful and therefore are dominant in our sample of actively traded distributions (evolutionary perspective). While information on the rationale behind rarity distributions is hard to retrieve \cite{punksoriginal}, the design and evolutionary explanations could have fuelled one another over time, with creators embedding rarity out of imitation of successful pre-existent collections. In this perspective, our results could help to further improve the design of NFT collections.

From the point of view of trading, it is important to highlight that our results concern genuinely emerging properties of the NFT market, since we only considered user-to-user sales. In doing so, we discarded the very first creator-to-user sales, which are often based on lotteries that prevent users to select what NFT to buy \cite{nft_mint_sale}. We found that while the impact of rarity is particularly strong for – and among – the rarest NFTs, which are thus genuinely non-fungible according to the market, its influence propagates to a large number of somehow rare NFTs (see Figure \ref{fig:impact_rarity_agg}g, inset and Figure \ref{fig:amplification_neg_return}a, inset). Most common NFTs, on the other hand, appear to behave more uniformly in the market, which appears to consider them essentially “fungible”. Overall, we anticipate that our results in this context may help inform the decisions of users interested in the financial aspects of NFTs.

Our study has limitations that future work could address.  First, our dataset is limited to collections available on Opensea, the biggest NFT market, and sold on the Ethereum blockchain. A natural extension would cover other platforms (potentially on other blockchains) and different types of NFTs, assessing whether rarity has the same effects on other kinds of NFTs such as those related to gaming and the metaverse. Second, we used the rarity score to quantify the rarity of an NFT. While this method does take into account every trait associated with an NFT, it does not consider possible combined effects stemming from the combination of multiple traits (e.g., two common traits for a collection might be present together in just one NFT, making it very rare). Future work could assess whether such second-order effects do play a role on the market performance of NFTs. Third, we considered traits as they are encoded in the NFT metadata and reported on rarity.tools, limiting the analysis to collections where such metadata are available and consistently recorded. Future work making use of computer vision techniques to extract human readable attributes from visual information of NFTs would yield to larger datasets and assess whether also less “official” visual traits, potentially shared by NFTs in multiple collections and where previously developed metrics might help~\cite{rabinowitz1981seven, violle2017functional}, might play a role on the NFT market. Finally, while this work has focused on how rarity affects NFT market success, a natural extension of the work should focus on how buyers behave with respect to rarity.

\section*{Acknowledgements}

The research was partly supported by The Alan Turing Institute. L.M.A. acknowledges the support from the Carlsberg Foundation through the COCOONS project (CF21-0432)

\section*{Data availability}

Data downloaded from the OpenSea API is available at \url{https://osf.io/7w9r6/}.

\section*{Author contributions}

A.M., A.Br., M.N., L.A., M.M., L.M.A., and A.Ba. designed the study. A.M. carried out the data collection. A.M. and A.Br. performed the measurements. A.M., A.Br., M.N., L.A., M.M., L.M.A., and A.Ba. analysed the data, discussed the results, and contributed to the final manuscript.

\section*{Competing interests}

The authors declare no competing interests.

\clearpage

\newpage

\begin{Large}{\begin{center}
{\bf{Supplementary Information for "Heterogeneous rarity patterns drive price dynamics in NFT collections"}}

\end{center} 
 }
\end{Large}
\thispagestyle{empty}
\tableofcontents
\clearpage

\appendix

\renewcommand{\figurename}{Supplementary Figure}

\setcounter{figure}{0}

\renewcommand{\tablename}{Supplementary Table}

\section{OpenSea market mechanisms}

For a majority of NFT collectibles, the minting happens as follows. The creators offer the possibility for anyone with a wallet to generate a new NFT for a fixed price, whose attributes will be randomly selected, even though each attribute can only be given to a specific amount of NFTs. Once every NFT has been minted by the community, they are made available to their buyers, who can sell them on a marketplace afterwards. 

Before releasing their collection, creators also set how much royalty they want to get from each secondary sale related to their NFTs. As such, every time a new sale happens, the royalty is deduced from the share the seller gets, as well at 2.5\% of the total price that OpenSea gets from every sale taking place on their platform. 

The following table details, for a few collections, the initial price at which the NFTs could be minted (gas fee, i.e., the fees required to conduct a transaction on the Ethereum blockchain, not included). Note that these transactions are not considered as sales per se by OpenSea's official API.

\begin{table}[H]
\centering
\begin{tabular}{|l|l|}
\hline
\textbf{Collection}   & \textbf{Minting Price}                                        \\ \hline
CryptoPunks           & Free                                                          \\ \hline
Bored Ape Yacht Club              & 0.08 ETH                                                          \\ \hline
World of Women              & 0.07 ETH                                                          \\ \hline
CryptoTrunks          & 0.5 ETH                                                       \\ \hline
CryptoCorgi           & First corgi to be claimed at 0.001 ETH, last one at 1.001 ETH \\ \hline
Sewer Rat Social Club & 0.05 ETH                                                      \\ \hline
Rabbit College Club   & 0.02 ETH                                                      \\ \hline
Cute Pig Club         & 0.03 ETH                                                      \\ \hline
Ape Gang              & Free                                                          \\ \hline

\end{tabular}
\end{table}

\section{Generative art mechanisms}
\label{section:gen_art}

As previously mentioned, NFT collectibles are usually generated using an algorithmic procedure, which can lead to thousands of unique tokens created with the same set of instructions \cite{generativeart}. However, the inner workings of the algorithms have not been shared by the creators, and can greatly differ between collections. It is therefore impossible to assess whether the rarity curves for the collections displayed in Section ``Quantifying Rarity'' share similarities because their algorithms follow similar steps. In the case of the CryptoPunks, members of the community have been attempting to reverse-engineer the algorithm used by Larva Labs to generate the original Punks \cite{punksoriginal}, or even to replicate it \cite{punksalgo, expansionpunks}. However, the creators never released any information on the matter, as well as any other NFT collectibles creator.

\section{List of collections}
\label{section:coll_list}

\begin{table}[H]
\centering
\scalebox{0.60}{
\begin{tabular}{|lllll|}
\hline
\multicolumn{5}{|c|}{\textbf{Collection Names}}                                                                                           \\ \hline
\multicolumn{1}{|l|}{0N1 Force}                      & \multicolumn{1}{l|}{0xVampire Project}                  & \multicolumn{1}{l|}{24px}                                       & \multicolumn{1}{l|}{8 BIT UNIVERSE}                & Absurd Arboretum                \\ \hline
\multicolumn{1}{|l|}{Adam Bomb Squad}                & \multicolumn{1}{l|}{AfroDroids By Owo}                  & \multicolumn{1}{l|}{Al Cabones}                                 & \multicolumn{1}{l|}{AlphaBetty Doodles}            & AmeegosOfficialNFT              \\ \hline
\multicolumn{1}{|l|}{Angels of Aether}               & \multicolumn{1}{l|}{Angry Boars}                        & \multicolumn{1}{l|}{AnimalWorldWar}                             & \multicolumn{1}{l|}{Animathereum}                  & Animetas                        \\ \hline
\multicolumn{1}{|l|}{Ape Gang}                       & \multicolumn{1}{l|}{Ape Harbour Yachts}                 & \multicolumn{1}{l|}{ApesOfSpace}                                & \multicolumn{1}{l|}{Approving Corgis}              & Arabian Camels                  \\ \hline
\multicolumn{1}{|l|}{ArcadeNFT}                      & \multicolumn{1}{l|}{Art Stars Club Official}            & \multicolumn{1}{l|}{Astro Frens}                                & \multicolumn{1}{l|}{Astrohedz}                     & Avarik Saga Universe            \\ \hline
\multicolumn{1}{|l|}{Avastars}                       & \multicolumn{1}{l|}{Axolittles}                         & \multicolumn{1}{l|}{BASTARD GAN PUNKS V2}                       & \multicolumn{1}{l|}{BLU Blox}                      & BULLSEUM                        \\ \hline
\multicolumn{1}{|l|}{BYOPills}                       & \multicolumn{1}{l|}{Baby Combat Bots G1}                & \multicolumn{1}{l|}{Bad Bunnies NFT}                            & \multicolumn{1}{l|}{Bad Kids Alley Official}       & Badass Bulls                    \\ \hline
\multicolumn{1}{|l|}{Barn Owls}                      & \multicolumn{1}{l|}{Barn Owls Dino Palz}                & \multicolumn{1}{l|}{Based Fish Mafia}                           & \multicolumn{1}{l|}{Bear Market Bears}             & Bears Deluxe                    \\ \hline
\multicolumn{1}{|l|}{BearsOnTheBlock}                & \multicolumn{1}{l|}{Beatnik Tiki Tribe}                 & \multicolumn{1}{l|}{Bit Wine}                                   & \multicolumn{1}{l|}{BlankFace}                     & Blob Mob                        \\ \hline
\multicolumn{1}{|l|}{BlockchainBikers}               & \multicolumn{1}{l|}{Bones \& Bananas}                   & \multicolumn{1}{l|}{Bones Club Heritage}                        & \multicolumn{1}{l|}{Bonsai by ZENFT}               & Bored Ape Kennel Club           \\ \hline
\multicolumn{1}{|l|}{Bored Ape Yacht Club}           & \multicolumn{1}{l|}{Bored Mummy Baby Waking Up}         & \multicolumn{1}{l|}{Bored Mummy Waking Up}                      & \multicolumn{1}{l|}{Boring Bananas Co.}            & Boss Beauties                   \\ \hline
\multicolumn{1}{|l|}{BroadcastersNFT}                & \multicolumn{1}{l|}{BullsOnTheBlock}                    & \multicolumn{1}{l|}{Bunker Beasts}                              & \multicolumn{1}{l|}{Buzzed Bear Hideout}           & CHIBI DINOS                     \\ \hline
\multicolumn{1}{|l|}{COVIDPunks!}                    & \multicolumn{1}{l|}{CanineCartel}                       & \multicolumn{1}{l|}{Cartlads}                                   & \multicolumn{1}{l|}{Catctus Collectibles}          & Catshit Crazy                   \\ \hline
\multicolumn{1}{|l|}{Chads NFT}                      & \multicolumn{1}{l|}{ChainFaces}                         & \multicolumn{1}{l|}{Chibi Apes}                                 & \multicolumn{1}{l|}{Chihuahua Gang}                & Chill Frogs NFT                 \\ \hline
\multicolumn{1}{|l|}{Chiptos}                        & \multicolumn{1}{l|}{Chubbies}                           & \multicolumn{1}{l|}{Ciphersquares Official}                     & \multicolumn{1}{l|}{Citizens of Bulliever Island}  & Claylings                       \\ \hline
\multicolumn{1}{|l|}{CleverGirls NFT}                & \multicolumn{1}{l|}{Cool Cats NFT}                      & \multicolumn{1}{l|}{Crazy Crows Chess Club}                     & \multicolumn{1}{l|}{Crazy Dragon Corps}            & Crazy Lizard Army               \\ \hline
\multicolumn{1}{|l|}{CrazySkullzNFT}                 & \multicolumn{1}{l|}{Criminal Donkeys}                   & \multicolumn{1}{l|}{Crumbys Bakery}                             & \multicolumn{1}{l|}{CrypToadz by GREMPLIN}         & Cryptinieis                     \\ \hline
\multicolumn{1}{|l|}{Crypto Cannabis Club}           & \multicolumn{1}{l|}{Crypto Corgis}                      & \multicolumn{1}{l|}{Crypto Duckies}                             & \multicolumn{1}{l|}{Crypto Ghosts NFT}             & Crypto Hobos                    \\ \hline
\multicolumn{1}{|l|}{Crypto Hodlers NFT}             & \multicolumn{1}{l|}{Crypto Squatches}                   & \multicolumn{1}{l|}{Crypto Tuners}                              & \multicolumn{1}{l|}{Crypto.Chicks}                 & CryptoFighters                  \\ \hline
\multicolumn{1}{|l|}{CryptoFinney}                   & \multicolumn{1}{l|}{CryptoMutts}                        & \multicolumn{1}{l|}{CryptoPunks}                                & \multicolumn{1}{l|}{CryptoSkulls}                  & CryptoTrunks                    \\ \hline
\multicolumn{1}{|l|}{Cunning Foxes}                  & \multicolumn{1}{l|}{Cupcats Official}                   & \multicolumn{1}{l|}{Cute Pig Club}                              & \multicolumn{1}{l|}{CyberKongz}                    & CyberKongz VX                   \\ \hline
\multicolumn{1}{|l|}{CyberPunkA12}                   & \multicolumn{1}{l|}{Cybergirl Fashion}                  & \multicolumn{1}{l|}{Cypher City}                                & \multicolumn{1}{l|}{Dapper Dinos Karma Collective} & Dapper Dinos NFT                \\ \hline
\multicolumn{1}{|l|}{Dapper Space Collective}        & \multicolumn{1}{l|}{Dead Devil Society}                 & \multicolumn{1}{l|}{DeadFellaz}                                 & \multicolumn{1}{l|}{DeadHeads}                     & Deadbears Official              \\ \hline
\multicolumn{1}{|l|}{Deez Nuts (Official Nuts)}      & \multicolumn{1}{l|}{Degen Gang}                         & \multicolumn{1}{l|}{Degenz}                                     & \multicolumn{1}{l|}{Delisted Tiny Punks}           & Derpy Birbs                     \\ \hline
\multicolumn{1}{|l|}{Devious Demon Dudes}            & \multicolumn{1}{l|}{Dizzy Dragons}                      & \multicolumn{1}{l|}{Doge Pound Puppies}                         & \multicolumn{1}{l|}{DogePirates}                   & Dogs Unchained                  \\ \hline
\multicolumn{1}{|l|}{Dope Shibas}                    & \multicolumn{1}{l|}{Dreamloops}                         & \multicolumn{1}{l|}{DystoPunks}                                 & \multicolumn{1}{l|}{Encryptas}                     & Epic Eagles                     \\ \hline
\multicolumn{1}{|l|}{Ether Cards Founder}            & \multicolumn{1}{l|}{EtherGals}                          & \multicolumn{1}{l|}{Ethereans Official}                         & \multicolumn{1}{l|}{Etheremura}                    & Evil Teddy Bear Club            \\ \hline
\multicolumn{1}{|l|}{FLUF World}                     & \multicolumn{1}{l|}{FUD Monsters}                       & \multicolumn{1}{l|}{FVCK\_CRYSTAL//}                            & \multicolumn{1}{l|}{FameLadySquas}                 & Fang Gang                       \\ \hline
\multicolumn{1}{|l|}{Garmers Marketverse Patrons}    & \multicolumn{1}{l|}{Fast Food Frens Collection}         & \multicolumn{1}{l|}{Fast Food Punks}                            & \multicolumn{1}{l|}{Fatales}                       & Flowtys                         \\ \hline
\multicolumn{1}{|l|}{Floyds World}                   & \multicolumn{1}{l|}{Forgotten Runes Wizards Cult}       & \multicolumn{1}{l|}{FoxyFam}                                    & \multicolumn{1}{l|}{Frogs In Disguise}             & FusionApes                      \\ \hline
\multicolumn{1}{|l|}{Fxck Face}                      & \multicolumn{1}{l|}{GLICPIXXXVER002 - GRAND COLLECTION} & \multicolumn{1}{l|}{GOATz}                                      & \multicolumn{1}{l|}{GRAYCRAFT2}                    & GRILLZ GANG                     \\ \hline
\multicolumn{1}{|l|}{Galactic Secret Agency}         & \multicolumn{1}{l|}{GalacticApes}                       & \multicolumn{1}{l|}{Galaxy Fight Club}                          & \multicolumn{1}{l|}{Galaxy-Eggs}                   & GameOfBlocks                    \\ \hline
\multicolumn{1}{|l|}{Gator World NFT}                & \multicolumn{1}{l|}{Gauntlets}                          & \multicolumn{1}{l|}{Genesis Block Art}                          & \multicolumn{1}{l|}{Glue Factory Show}             & Goblin Goons                    \\ \hline
\multicolumn{1}{|l|}{Good Guys NFT}                  & \multicolumn{1}{l|}{Goons of Balatroon}                 & \multicolumn{1}{l|}{Gorilla Nemesis}                            & \multicolumn{1}{l|}{Great Ape Society}             & Guardians of the Metaverse      \\ \hline
\multicolumn{1}{|l|}{Gutter Cat Gang}                & \multicolumn{1}{l|}{Gutter Rats}                        & \multicolumn{1}{l|}{HDPunks}                                    & \multicolumn{1}{l|}{HODL GANG}                     & Hammys                          \\ \hline
\multicolumn{1}{|l|}{HappyLand Gummy Bears Official} & \multicolumn{1}{l|}{HashGuise Gen One}                  & \multicolumn{1}{l|}{Hashmasks}                                  & \multicolumn{1}{l|}{HatchDracoNFT}                 & Heroes of Evermore              \\ \hline
\multicolumn{1}{|l|}{Hewer Clan}                     & \multicolumn{1}{l|}{HodlHeads}                          & \multicolumn{1}{l|}{Holy Cows}                                  & \multicolumn{1}{l|}{HypeHippo.io}                  & IMMORTALZ - Ambary Assassins    \\ \hline
\multicolumn{1}{|l|}{Incognito}                      & \multicolumn{1}{l|}{Kamagang}                           & \multicolumn{1}{l|}{Keplers Civil Society}                      & \multicolumn{1}{l|}{KidPunks}                      & Knights of Degen - Knights!     \\ \hline
\multicolumn{1}{|l|}{Koala Intelligence Agency}      & \multicolumn{1}{l|}{Koin Games Dev Squad}               & \multicolumn{1}{l|}{Kokeshi World}                              & \multicolumn{1}{l|}{Krazy Koalas NFT}              & Lamb Duhs                       \\ \hline
\multicolumn{1}{|l|}{Lazy Lions}                     & \multicolumn{1}{l|}{Lazy Lions Bungalows}               & \multicolumn{1}{l|}{Lobby Lobsters}                             & \multicolumn{1}{l|}{Lockdown Lemmings}             & Lonely Planet Space Observatory \\ \hline
\multicolumn{1}{|l|}{Long Neckie Fellas}             & \multicolumn{1}{l|}{Long Neckie Ladies}                 & \multicolumn{1}{l|}{Loopy Donuty}                               & \multicolumn{1}{l|}{Loot (for Adventurers)}        & Lost Souls Sanctuary            \\ \hline
\end{tabular}}
\end{table}

\begin{table}[H]
\centering
\scalebox{0.60}{
\begin{tabular}{|lllll|}
\hline
\multicolumn{5}{|c|}{\textbf{Collection Names}}                                                                                           \\ \hline
\multicolumn{1}{|l|}{Lostboy NFT}                    & \multicolumn{1}{l|}{Lucha Libre Knockout}               & \multicolumn{1}{l|}{Lucky Maneki}                               & \multicolumn{1}{l|}{Lucky Sloths NFT}              & Lumps World                     \\ \hline
\multicolumn{1}{|l|}{Lysergic Labs Shroomz}          & \multicolumn{1}{l|}{MOONDOGS ODYSSEY}                   & \multicolumn{1}{l|}{Mad Banana Union}                           & \multicolumn{1}{l|}{Mad Cat Militia}               & MaestroPups                     \\ \hline
\multicolumn{1}{|l|}{Magic Mushroom Clubhouse}       & \multicolumn{1}{l|}{Mandelbrot Set Collection}          & \multicolumn{1}{l|}{Maneki Gang}                                & \multicolumn{1}{l|}{MarsCatsVoyage}                & Meebits                         \\ \hline
\multicolumn{1}{|l|}{Mighty Manateez}                & \multicolumn{1}{l|}{Mini Monkey Mafia}                  & \multicolumn{1}{l|}{Minimints}                                  & \multicolumn{1}{l|}{MissCryptoClub}                & MjiBots                         \\ \hline
\multicolumn{1}{|l|}{Monas}                          & \multicolumn{1}{l|}{MonkePunks}                         & \multicolumn{1}{l|}{Monkeybrix}                                 & \multicolumn{1}{l|}{Monster Blocks - Official}     & Monster Rehab 1.0               \\ \hline
\multicolumn{1}{|l|}{Mutant Ape Yacht Club}          & \multicolumn{1}{l|}{MutantKongz}                        & \multicolumn{1}{l|}{Muttinks}                                   & \multicolumn{1}{l|}{My Fucking Pickle}             & NFT Siblings                    \\ \hline
\multicolumn{1}{|l|}{NFTBOY: Bored Ape Racers}       & \multicolumn{1}{l|}{NOOBS NFT}                          & \multicolumn{1}{l|}{Naughty Tigers Costume Club}                & \multicolumn{1}{l|}{Neon Junkies}                  & Nice Drips                      \\ \hline
\multicolumn{1}{|l|}{Nifty League DEGENs}            & \multicolumn{1}{l|}{Niftyriots}                         & \multicolumn{1}{l|}{Non-Fungible Heroes}                        & \multicolumn{1}{l|}{Notorious Frogs}               & ORCZ!                           \\ \hline
\multicolumn{1}{|l|}{OctoHedz}                       & \multicolumn{1}{l|}{Oddball Club (Official)}            & \multicolumn{1}{l|}{Official DogeX}                             & \multicolumn{1}{l|}{Omnimorphs}                    & OnChainMonkey                   \\ \hline
\multicolumn{1}{|l|}{Osiris Cosmic Kids}             & \multicolumn{1}{l|}{PEACEFUL GROUPIES}                  & \multicolumn{1}{l|}{PORK1984}                                   & \multicolumn{1}{l|}{POW NFT}                       & PPPandas                        \\ \hline
\multicolumn{1}{|l|}{Paladin Pandas}                 & \multicolumn{1}{l|}{Panda Dynasty}                      & \multicolumn{1}{l|}{Panda Golf Squad}                           & \multicolumn{1}{l|}{Party Penguins}                & Penguin Fight Club              \\ \hline
\multicolumn{1}{|l|}{PinapplesDayOut}                & \multicolumn{1}{l|}{Pirate Treasure Booty Club}         & \multicolumn{1}{l|}{PixaWizards}                                & \multicolumn{1}{l|}{Platy Punks - Official}        & PogPunks NFT                    \\ \hline
\multicolumn{1}{|l|}{Polar Pals Bobsledding}         & \multicolumn{1}{l|}{Posh Pandas}                        & \multicolumn{1}{l|}{Potato Power Club}                          & \multicolumn{1}{l|}{Primate Social Society}        & Procedural Space                \\ \hline
\multicolumn{1}{|l|}{Pudgy Penguins}                 & \multicolumn{1}{l|}{PunkBabies}                         & \multicolumn{1}{l|}{PunkCats}                                   & \multicolumn{1}{l|}{PunkScapes}                    & Purrnelopes Country Club        \\ \hline
\multicolumn{1}{|l|}{PyMons}                         & \multicolumn{1}{l|}{Qubits On The Ice}                  & \multicolumn{1}{l|}{RUUMZ}                                      & \multicolumn{1}{l|}{Rabbit College Club}           & Raccoon Mafia                   \\ \hline
\multicolumn{1}{|l|}{Raccoons Club}                  & \multicolumn{1}{l|}{RagingRhinos}                       & \multicolumn{1}{l|}{Re-Genz}                                    & \multicolumn{1}{l|}{Ready Player Cat NFT}          & Reb3l Bots                      \\ \hline
\multicolumn{1}{|l|}{Reckless Whales}                & \multicolumn{1}{l|}{RichKidsOfficial}                   & \multicolumn{1}{l|}{Rickstro Frens}                             & \multicolumn{1}{l|}{Rivermen}                      & Roaring Leaders                 \\ \hline
\multicolumn{1}{|l|}{Robotos Offocial}               & \multicolumn{1}{l|}{Rogue Society Bots}                 & \multicolumn{1}{l|}{Royal Ceramic Club}                         & \multicolumn{1}{l|}{Royal Society Chips}           & Royal Society of Players        \\ \hline
\multicolumn{1}{|l|}{Rumble Kong League}             & \multicolumn{1}{l|}{SLOTHz}                             & \multicolumn{1}{l|}{STRAWBERRY.WTF}                             & \multicolumn{1}{l|}{SVINS}                         & Sad Frogs District              \\ \hline
\multicolumn{1}{|l|}{Sad Girls Bar}                  & \multicolumn{1}{l|}{SamuraiDoge}                        & \multicolumn{1}{l|}{Sappy Seals}                                & \multicolumn{1}{l|}{Satoshibles}                   & Savage Droids                   \\ \hline
\multicolumn{1}{|l|}{Save the Martians}              & \multicolumn{1}{l|}{ScoopDog Squad}                     & \multicolumn{1}{l|}{Secret Society of Whales}                   & \multicolumn{1}{l|}{Sewer Rat Social Club}         & Shabu Town Shibas               \\ \hline
\multicolumn{1}{|l|}{Shaggy Sheep}                   & \multicolumn{1}{l|}{Shiba Society}                      & \multicolumn{1}{l|}{Sidus NFT Heroes}                           & \multicolumn{1}{l|}{SingularityHeroes}             & Sipherian Surge                 \\ \hline
\multicolumn{1}{|l|}{Skvullpvnks Hideout}            & \multicolumn{1}{l|}{Slacker Duck Pond}                  & \multicolumn{1}{l|}{Sleeper Hits Collection Volume 1 NFT Cribs} & \multicolumn{1}{l|}{Slimes World}                  & Slumdoge Billionaires           \\ \hline
\multicolumn{1}{|l|}{Sneaky Vampire Syndicate}       & \multicolumn{1}{l|}{Soccer Doge Club}                   & \multicolumn{1}{l|}{Space Dinos Club}                           & \multicolumn{1}{l|}{Space Poggers}                 & SpacePunksClub                  \\ \hline
\multicolumn{1}{|l|}{SpaceShibas}                    & \multicolumn{1}{l|}{Spookies NFT}                       & \multicolumn{1}{l|}{SportsIcon Lion Club}                       & \multicolumn{1}{l|}{Spunks}                        & Standametti                     \\ \hline
\multicolumn{1}{|l|}{Stoned Apez Saturn Club}        & \multicolumn{1}{l|}{Stoner Cats}                        & \multicolumn{1}{l|}{Stranger EggZ}                              & \multicolumn{1}{l|}{StripperVille NFTs}            & SupDucks                        \\ \hline
\multicolumn{1}{|l|}{Super Yeti}                     & \multicolumn{1}{l|}{Superfuzz The Bad Batch}            & \multicolumn{1}{l|}{Superfuzz The Good Guys}                    & \multicolumn{1}{l|}{Sushiverse}                    & SympathyForTheDevils            \\ \hline
\multicolumn{1}{|l|}{THE PLUTO ALLIANCE}             & \multicolumn{1}{l|}{THE SHRUNKENHEADZ}                  & \multicolumn{1}{l|}{The Alien Boy}                              & \multicolumn{1}{l|}{The BirdHouse}                 & The CryptoDads                  \\ \hline
\multicolumn{1}{|l|}{The CryptoSaints}               & \multicolumn{1}{l|}{The Doge Pound}                     & \multicolumn{1}{l|}{The Fuckin' Trolls}                         & \multicolumn{1}{l|}{The Goobers}                   & The Graveyard Sale              \\ \hline
\multicolumn{1}{|l|}{The KILLAZ}                     & \multicolumn{1}{l|}{The KittyButts}                     & \multicolumn{1}{l|}{The League Of Sacred Devils}                & \multicolumn{1}{l|}{The Lost Glitches}             & The MonstroCities               \\ \hline
\multicolumn{1}{|l|}{The Moon Boyz}                  & \multicolumn{1}{l|}{The NFTBirds}                       & \multicolumn{1}{l|}{The Nanoz}                                  & \multicolumn{1}{l|}{The Nemesis Companions}        & The Ninja Hideout               \\ \hline
\multicolumn{1}{|l|}{The Project URS}                & \multicolumn{1}{l|}{The Sevens (Official)}              & \multicolumn{1}{l|}{The Shark Cove}                             & \multicolumn{1}{l|}{The Soldiers Of The Metaverse} & The Street Dawgs                \\ \hline
\multicolumn{1}{|l|}{The Unstable Horses Yard}       & \multicolumn{1}{l|}{The Vogu Collective}                & \multicolumn{1}{l|}{The Wanderers}                              & \multicolumn{1}{l|}{The Wicked Craniums}           & The Wicked Stallions            \\ \hline
\multicolumn{1}{|l|}{The WolfGang Pups}              & \multicolumn{1}{l|}{The WonderQuest}                    & \multicolumn{1}{l|}{The WynLambo}                               & \multicolumn{1}{l|}{TheHeartProject}               & TheTigersGuild                  \\ \hline
\multicolumn{1}{|l|}{Tie Dye Ninjas}                 & \multicolumn{1}{l|}{Tokenmon}                           & \multicolumn{1}{l|}{Tools of Rock}                              & \multicolumn{1}{l|}{Top Dog Beach Club}            & TradeSquads                     \\ \hline
\multicolumn{1}{|l|}{Trollz}                         & \multicolumn{1}{l|}{Ugly Cuties Art Club (UCAC)}        & \multicolumn{1}{l|}{United Punks Union}                         & \multicolumn{1}{l|}{Untamed Elephants}             & Unusual Whales                  \\ \hline
\multicolumn{1}{|l|}{VeeFriends}                     & \multicolumn{1}{l|}{Vegiemon}                           & \multicolumn{1}{l|}{Vox Collectibles}                           & \multicolumn{1}{l|}{Voxies}                        & WE ARE THE OUTKAST              \\ \hline
\multicolumn{1}{|l|}{Waifusion}                      & \multicolumn{1}{l|}{Wall Street Chads}                  & \multicolumn{1}{l|}{Wanna Panda}                                & \multicolumn{1}{l|}{Wannabes Music Club}           & Warriors of Aradena             \\ \hline
\multicolumn{1}{|l|}{We are Dorkis}                  & \multicolumn{1}{l|}{WeMint Washington}                  & \multicolumn{1}{l|}{Weird Whales}                               & \multicolumn{1}{l|}{Wicked Ape Bone Club}          & Wicked Hound Bone Club          \\ \hline
\multicolumn{1}{|l|}{Wild Stag Treehouse}            & \multicolumn{1}{l|}{Winter Bears}                       & \multicolumn{1}{l|}{Woodies Generative Characters}              & \multicolumn{1}{l|}{World of Women}                & Zunks                           \\ \hline
\multicolumn{1}{|l|}{astroGems}                      & \multicolumn{1}{l|}{bastard gan penguins}               & \multicolumn{1}{l|}{isotile Genesis Avatars}                    & \multicolumn{1}{l|}{thedudes}                      & uwucrew                         \\ \hline

\end{tabular}}
\end{table}

\section{Rarity score distributions}

As detailed in the main text, we use Akaike Information Criterium~\cite{wagenmakers2004aic} and Maximum Likelihood Estimation to determine the distribution that best describes the rarity score distribution for each collection. 
We select the distribution among a subset of distributions implemented in the scipy.stats python package, requiring the distributions to be heterogeneous, continuous, and with at most 3 parameters (including location and scale).
This results in choosing among the following distributions: \textit{uniform}, \textit{pareto}, \textit{cauchy}, \textit{lognormal}, \textit{levy}, \textit{exponential}.

\section{Supplementary figures}

\begin{figure}[!htb]
    \centering
    \includegraphics[width=0.9\textwidth]{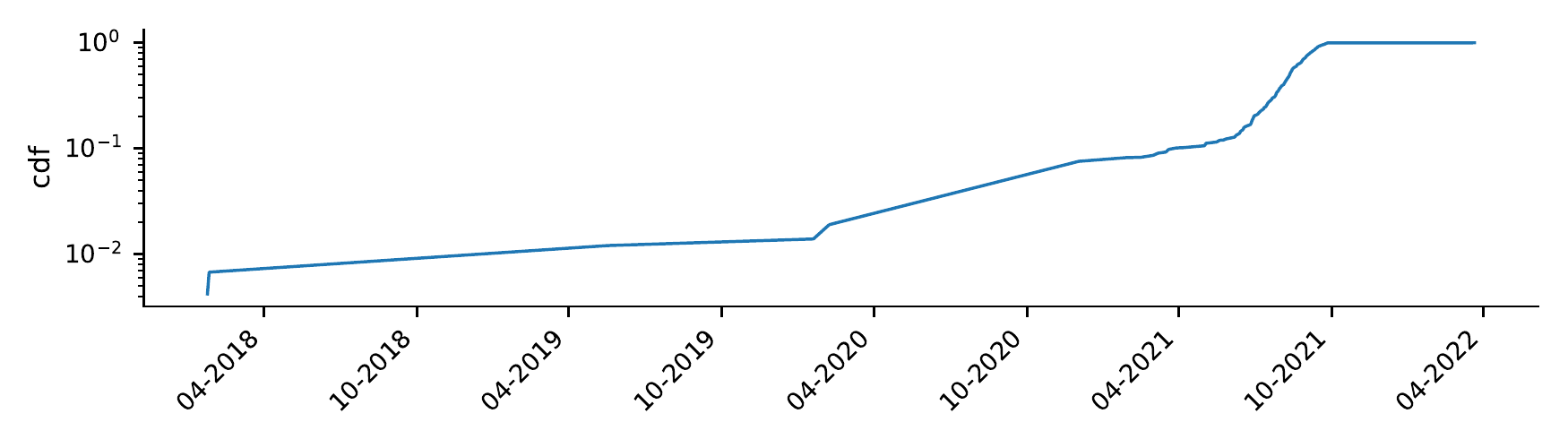}\\
    \caption{\textbf{Collectible NFTs minted over time.} Distribution of the collectible NFTs considered in this analysis minted over time.}
    \label{fig:minted_nfts}
\end{figure}

\begin{figure}[!htb]
    \centering
    \includegraphics[width=0.9\textwidth]{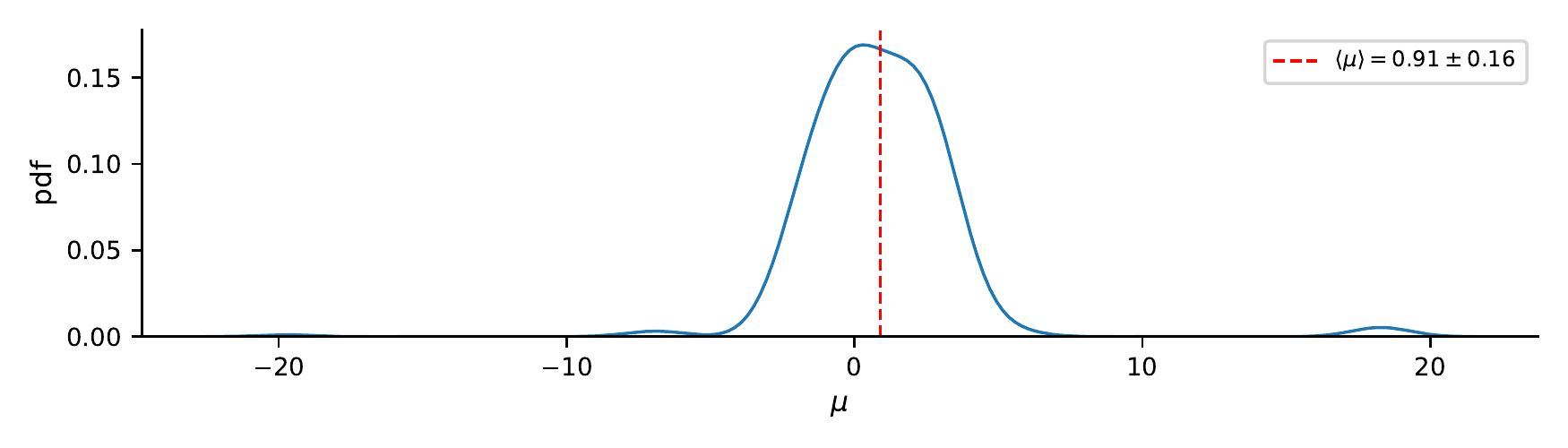}\\
    \caption{\textbf{Distribution of the log-normal distribution characteristic parameter $\mu$.} Distribution of the log-normal distribution parameter $\mu$ (blue line), and its average value across collections (red dashed line). The log-normal distribution $\ln(X)\sim {\mathcal{N}}(\mu, \sigma^{2})$ captures the distribution of rarity for $90\%$ of collections.}
    \label{fig:mean_lognorm}
\end{figure}

\begin{figure}[!htb]
    \centering
    \includegraphics[width=0.99\textwidth]{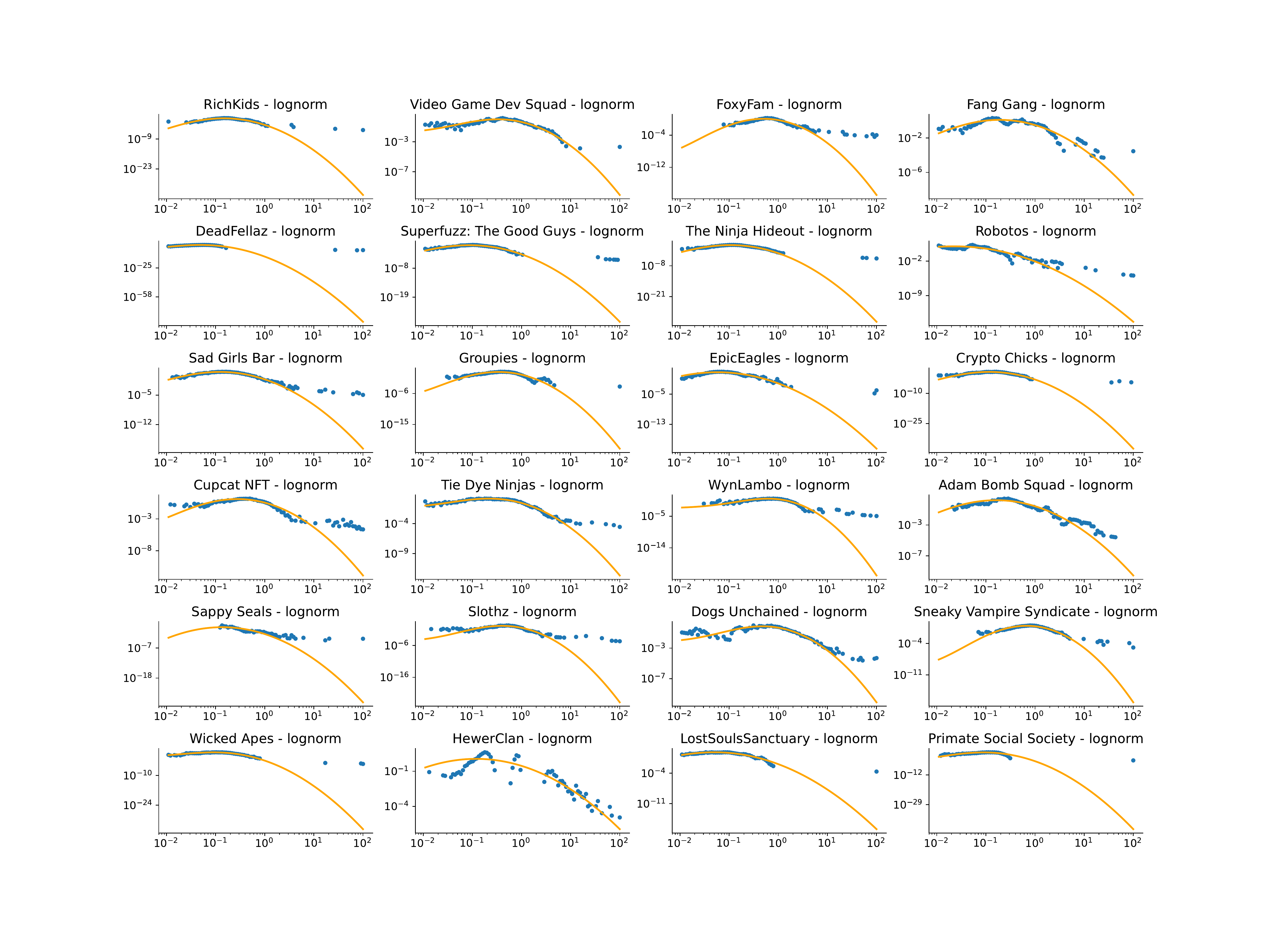}\\
    \caption{\textbf{Fitting of the rarity distribution.} Distribution of the rarity score of the NFTs within several collections included in the dataset (blue dots), along with the best distribution fit computed using Maximum Likelihood Estimation and Akaike Information Criterium~\cite{wagenmakers2004aic} (orange line).}
    \label{fig:best_fits}
\end{figure}

\clearpage
\subsection{Rarity rank results}

\begin{figure}[!htb]
    \centering
    \includegraphics[width=0.9\textwidth]{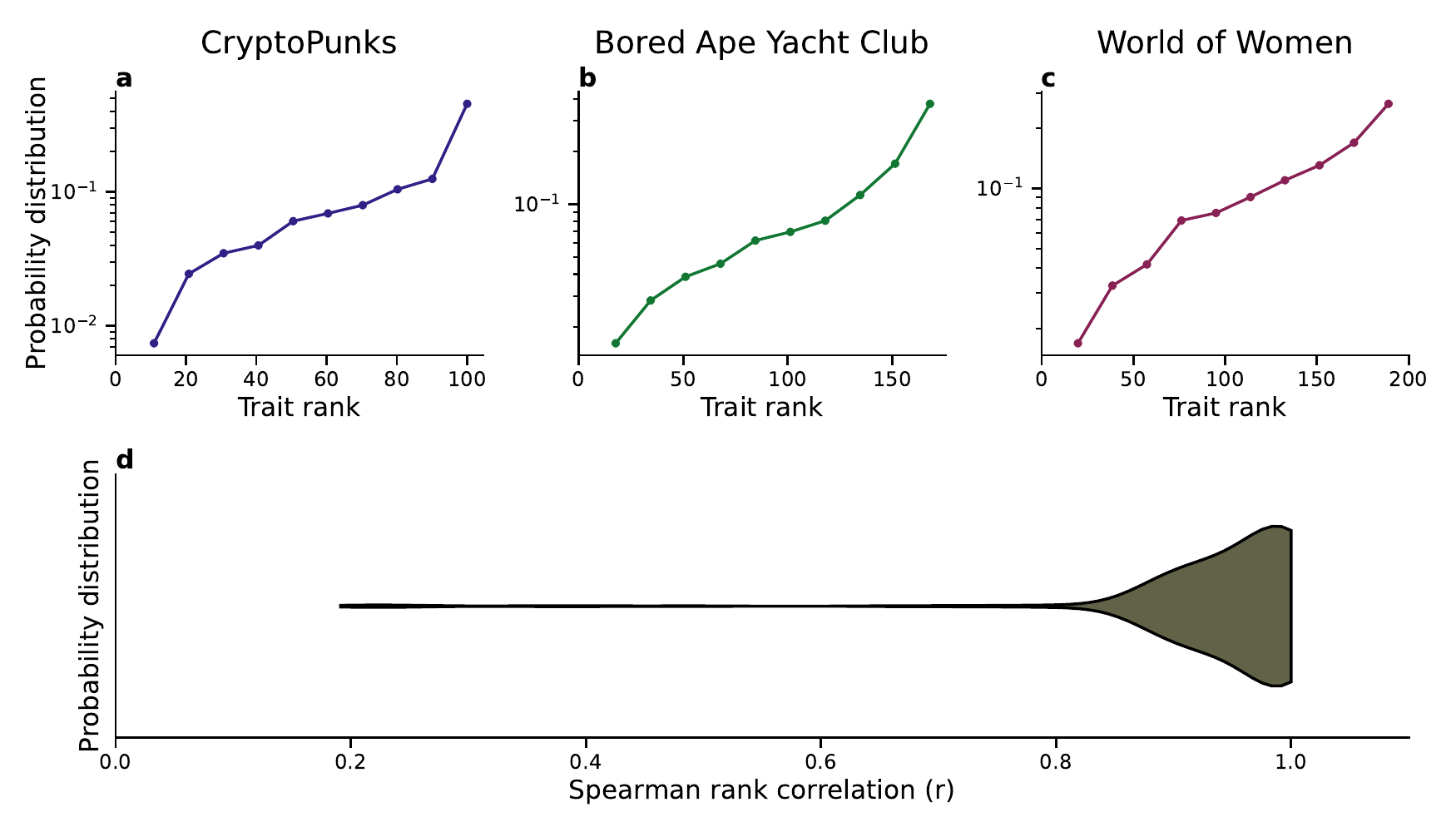}\\
    \caption{\textbf{Trait Rarity Rank.} 
    a-c) Distribution of the trait rarity rank of the NFTs within three collections: CryptoPunks (a), Bored Ape Yacht Club (b), and World of Women (c). d) Violin plot of the Spearman Rank correlation computed between the rarity rank and the number of NFTs with that rank.}
    \label{fig:rarity_design_rank}
\end{figure}

\begin{figure}[!htb]
    \centering
    \includegraphics[width=0.9\textwidth]{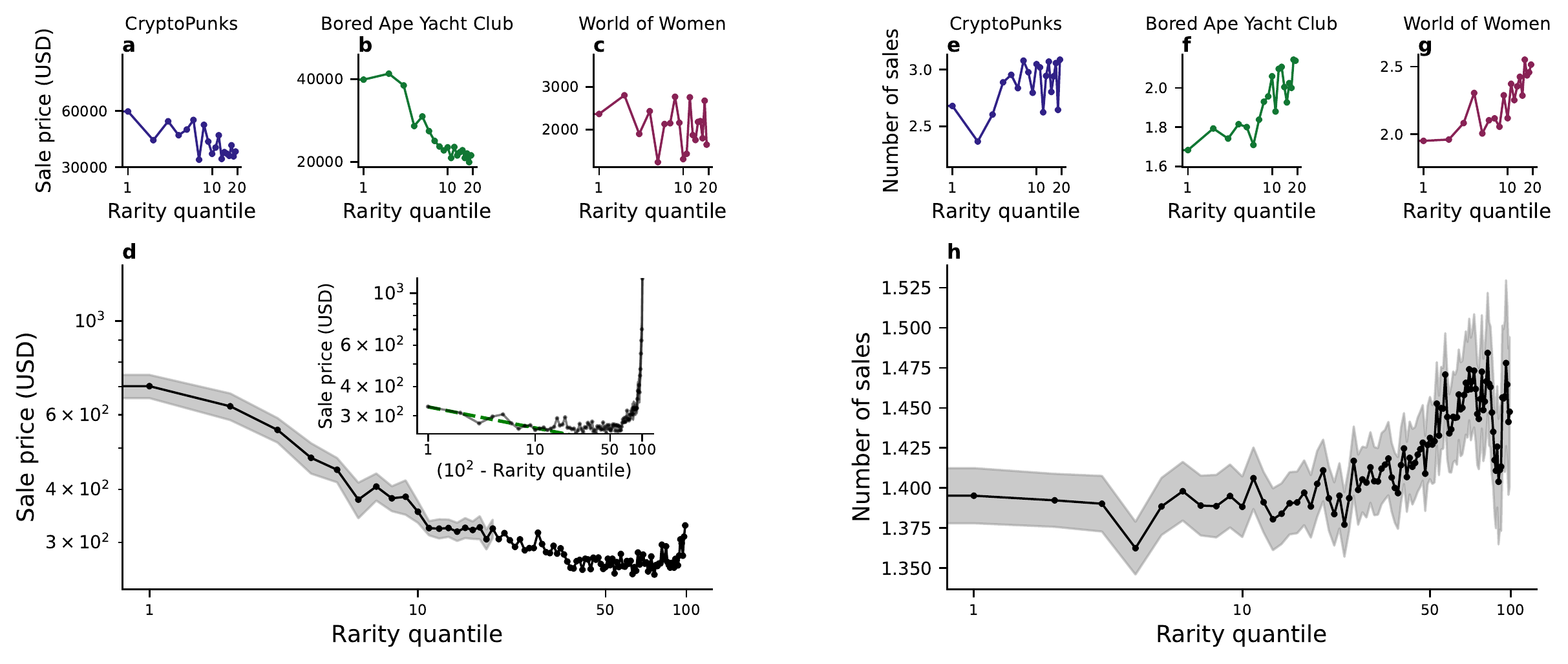}\\
    \caption{\textbf{Rare NFTs have a higher financial value and circulate less on the marketplace - Analysis with the rarity rank.} 
     Median sale price in USD (a-c) and average number of sales (e-g) by rarity quantile (with 20 quantiles considered) for three collections: CryptoPunks (a and e), Bored Ape Yacht Club (b and f), and World Women (c and g). d) Median sale price by rarity quantile (with 100 quantiles considered) considering all collections. Inset: median sale price against the quantity (100-$q$), where $q$ is the rarity quantile, in log-log scale (black line) and the corresponding power law fit (green dashed line). h) Median number of sales by rarity quantile considering all collections. The NFTs are aggregated by quantile depending on their rarity rank, i.e the first quantile represents the rarest NFTs within the collection.}
    \label{fig:impact_coll_trait}
\end{figure}

\begin{figure}[!htb]
    \centering
    \includegraphics[width=0.9\textwidth]{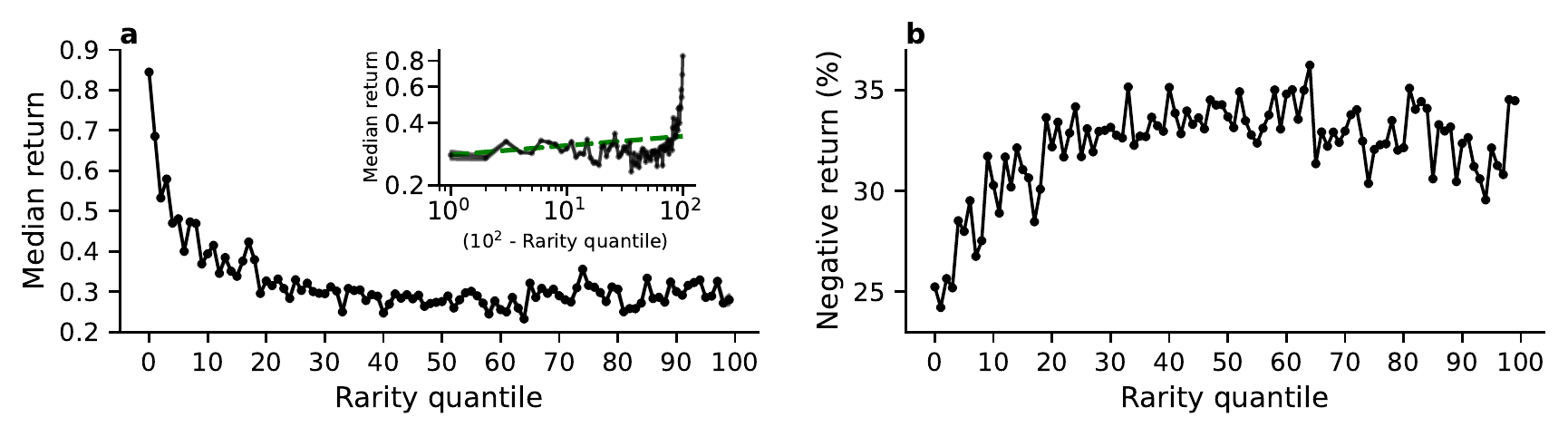}\\
    \caption{\textbf{High rarity leads to higher returns, and a lower chance of a negative return - Analysis with the rarity rank.} a) Median return in USD by rarity quantile. Inset: median return against the quantity (100-$q$), where $q$ is the rarity quantile in log-log scale (black line) and the corresponding power law fit (green dashed line). b) Fraction of sales with negative return in USD by rarity quantile. The NFTs are aggregated by quantile depending on their rarity rank, i.e the first quantile represents the rarest NFTs within the collection.
    }
    \label{fig:amplification_neg_return_trait}
\end{figure}

\clearpage
\subsection{Currency robustness check}

\begin{figure}[!htb]
    \centering
    \includegraphics[width=0.9\textwidth]{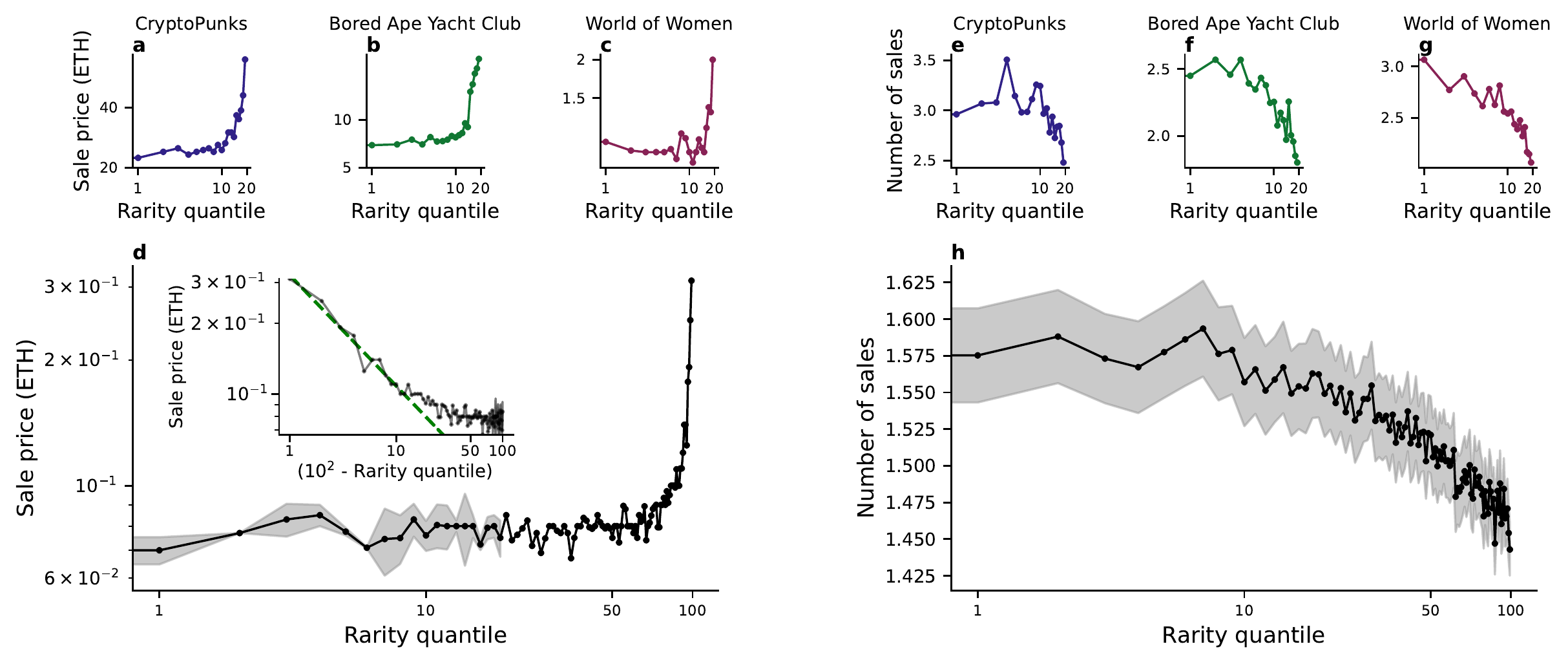}\\
    \caption{\textbf{Rare NFTs have a higher financial value and circulate less on the marketplace - Price in ETH.} 
     Median sale price in ETH (a-c) and average number of sales (e-g) by rarity quantile (with 20 quantiles considered) for three collections: CryptoPunks (a and e), Bored Ape Yacht Club (b and f), and World Women (c and g). d) Median sale price by rarity quantile (with 100 quantiles considered) considering all collections. Inset: median sale price against the quantity (100-$q$), where $q$ is the rarity quantile, in log-log scale (black line) and the corresponding power law fit (green dashed line). h) Median number of sales by rarity quantile considering all collections.}
    \label{fig:impact_coll_eth}
\end{figure}

\begin{figure}[!htb]
    \centering
    \includegraphics[width=0.9\textwidth]{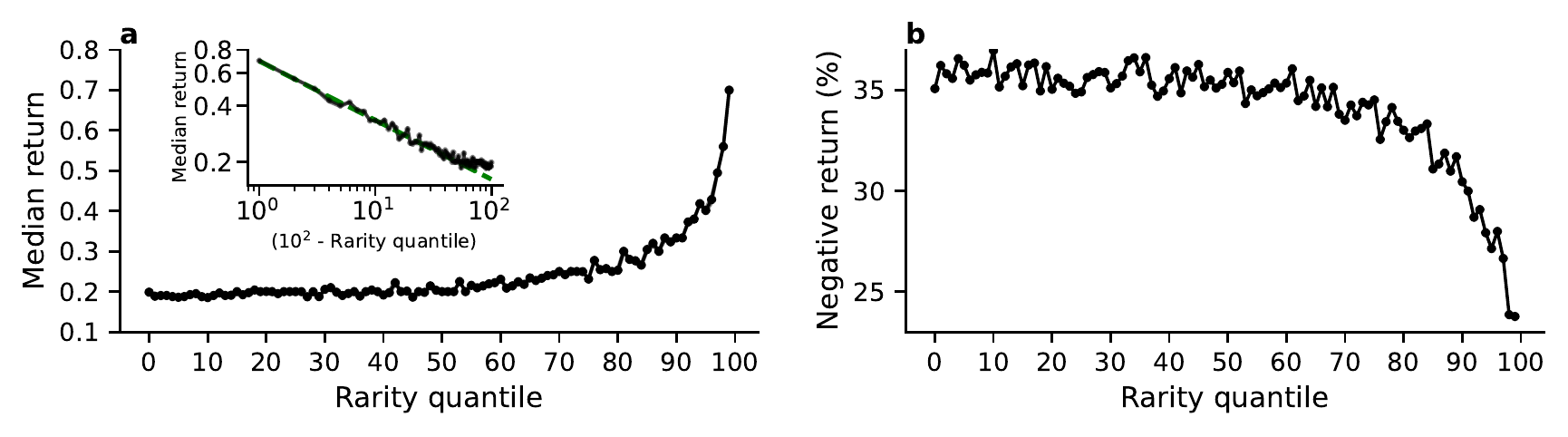}\\
    \caption{\textbf{High rarity leads to higher returns, and a lower chance of a negative return - Price in ETH.} a) Median return in ETH by rarity quantile. Inset: median return against the quantity (100-$q$), where $q$ is the rarity quantile in log-log scale (black line) and the corresponding power law fit (green dashed line). b) Fraction of sales with negative return in ETH by rarity quantile.}
    \label{fig:amplification_neg_return_eth}
\end{figure}

\clearpage
\subsection{Time robustness check}

To make sure that the findings we highlight in this paper are time-independent, we ran the same analysis by using only the transactions happening during specific time periods, to see whether we observe the same mechanisms within the marketplace. Therefore, we performed the analysis on the two last quarters of 2021, i.e., first on Q3 (July - September 2021) and then on Q4 (October - December 2021).

\begin{figure}[!htb]
    \centering
    \includegraphics[width=0.9\textwidth]{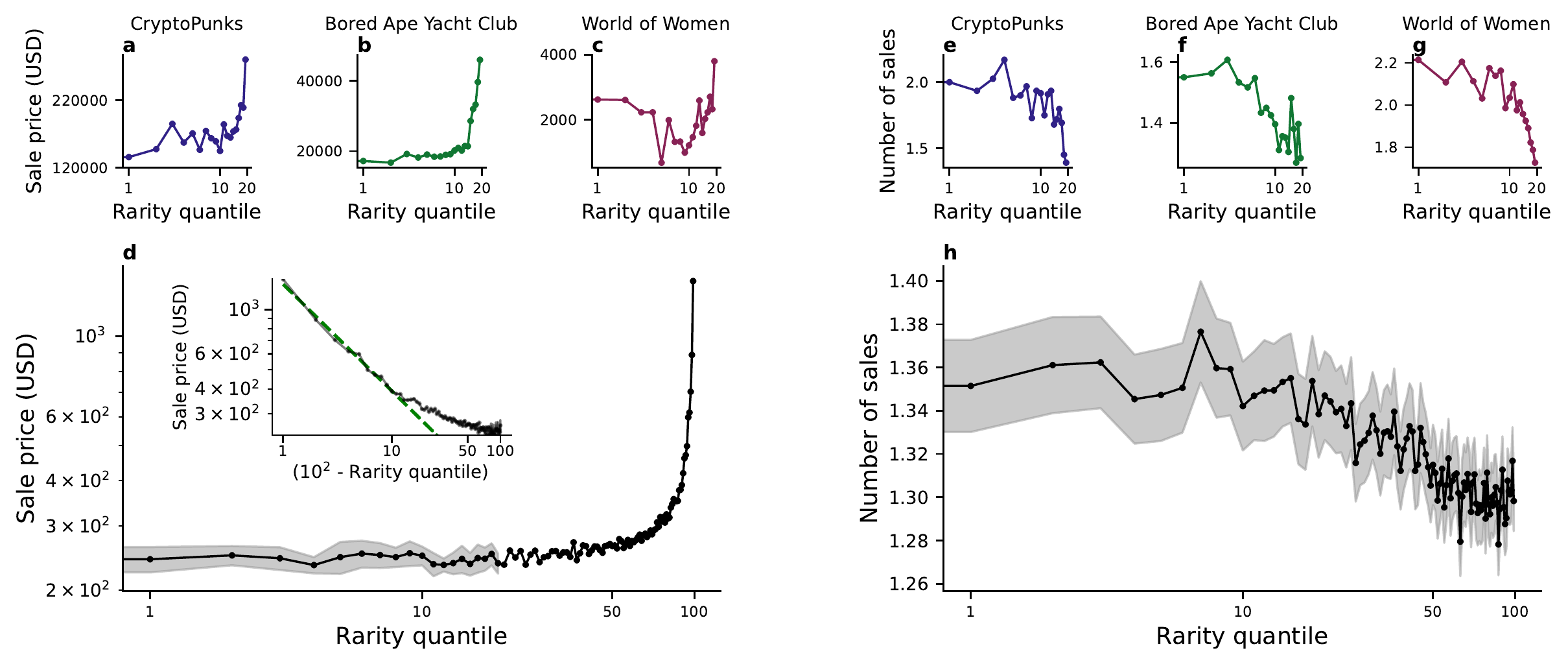}\\
    \caption{\textbf{Rare NFTs have a higher financial value and circulate less on the marketplace - Analysis on Q3 2021.} 
     Median sale price in USD (a-c) and average number of sales (e-g) by rarity quantile (with 20 quantiles considered) for three collections: CryptoPunks (a and e), Bored Ape Yacht Club (b and f), and World Women (c and g). d) Median sale price by rarity quantile (with 100 quantiles considered) considering all collections. Inset: median sale price against the quantity (100-$q$), where $q$ is the rarity quantile, in log-log scale (black line) and the corresponding power law fit (green dashed line). h) Median number of sales by rarity quantile considering all collections. This analysis only takes into consideration the sales happening during Q3 2021.}
    \label{fig:impact_coll_q3}
\end{figure}

\begin{figure}[!htb]
    \centering
    \includegraphics[width=0.9\textwidth]{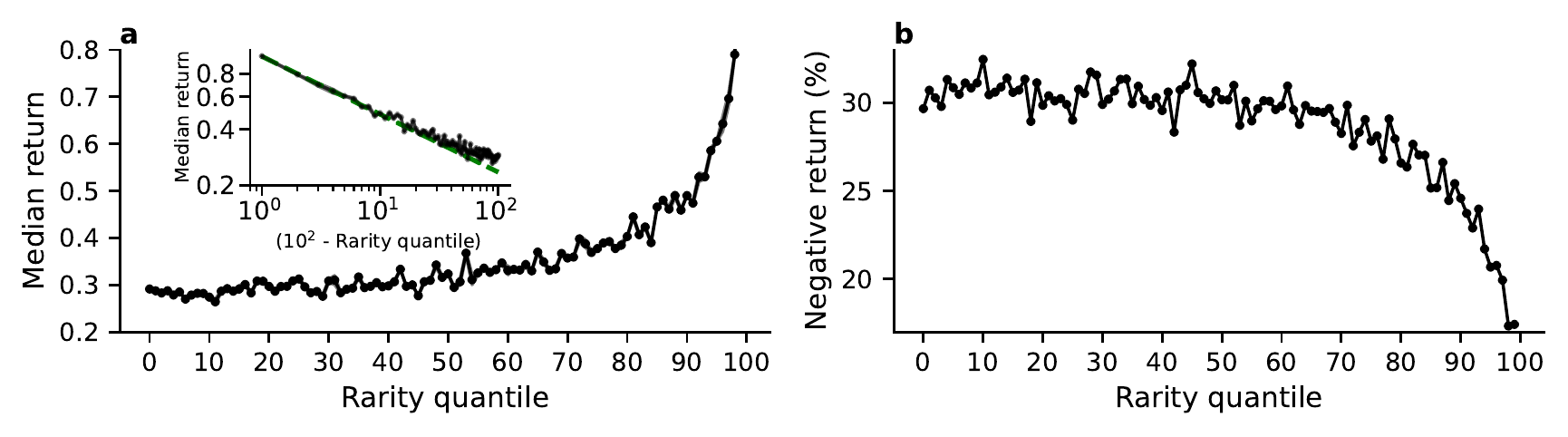}\\
    \caption{\textbf{High rarity leads to higher returns, and a lower chance of a negative return - Analysis on Q3 2021.} a) Median return in USD by rarity quantile. Inset: median return against the quantity (100-$q$), where $q$ is the rarity quantile in log-log scale (black line) and the corresponding power law fit (green dashed line). b) Fraction of sales with negative return in USD by rarity quantile. This analysis only takes into consideration the sales happening during Q3 2021.
    }
    \label{fig:amplification_neg_return_q3}
\end{figure}

\begin{figure}[!htb]
    \centering
    \includegraphics[width=0.9\textwidth]{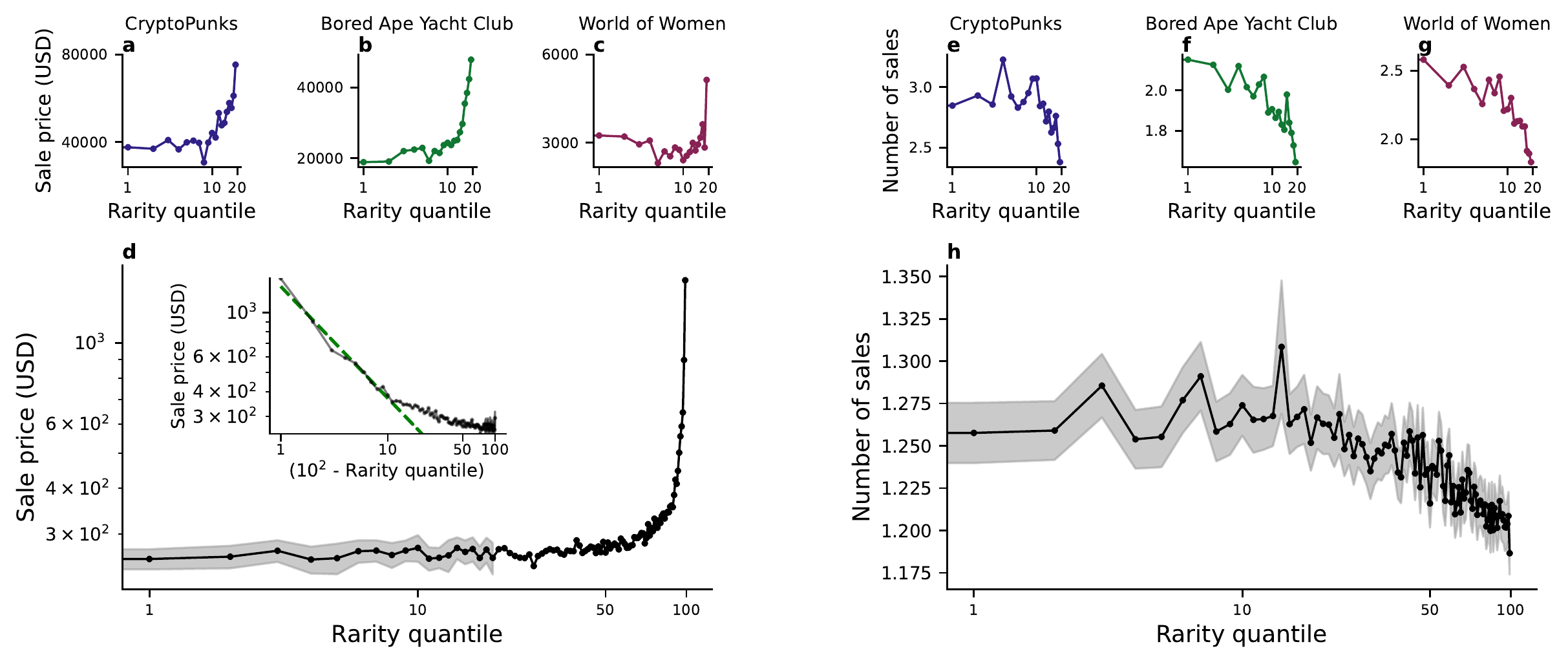}\\
    \caption{\textbf{Rare NFTs have a higher financial value and circulate less on the marketplace - Analysis on Q4 2021.} 
     Median sale price in USD (a-c) and average number of sales (e-g) by rarity quantile (with 20 quantiles considered) for three collections: CryptoPunks (a and e), Bored Ape Yacht Club (b and f), and World Women (c and g). d) Median sale price by rarity quantile (with 100 quantiles considered) considering all collections. Inset: median sale price against the quantity (100-$q$), where $q$ is the rarity quantile, in log-log scale (black line) and the corresponding power law fit (green dashed line). h) Median number of sales by rarity quantile considering all collections. This analysis only takes into consideration the sales happening during Q4 2021.}
    \label{fig:impact_coll_q4}
\end{figure}

\begin{figure}[!htb]
    \centering
    \includegraphics[width=0.9\textwidth]{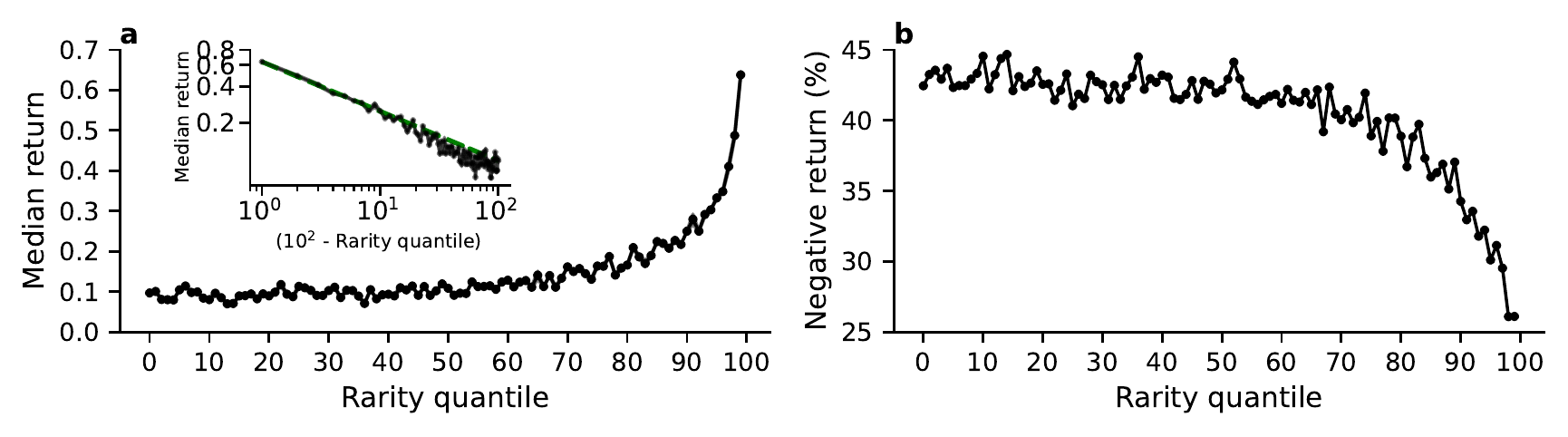}\\
    \caption{\textbf{High rarity leads to higher returns, and a lower chance of a negative return - Analysis on Q4 2021.} a) Median return in USD by rarity quantile. Inset: median return against the quantity (100-$q$), where $q$ is the rarity quantile in log-log scale (black line) and the corresponding power law fit (green dashed line). b) Fraction of sales with negative return in USD by rarity quantile. This analysis only takes into consideration the sales happening during Q4 2021.
    }
    \label{fig:amplification_neg_return_q4}
\end{figure}

\clearpage
\subsection{Tails robustness check}

\begin{figure}[!htb]
    \centering
    \includegraphics[width=1\textwidth]{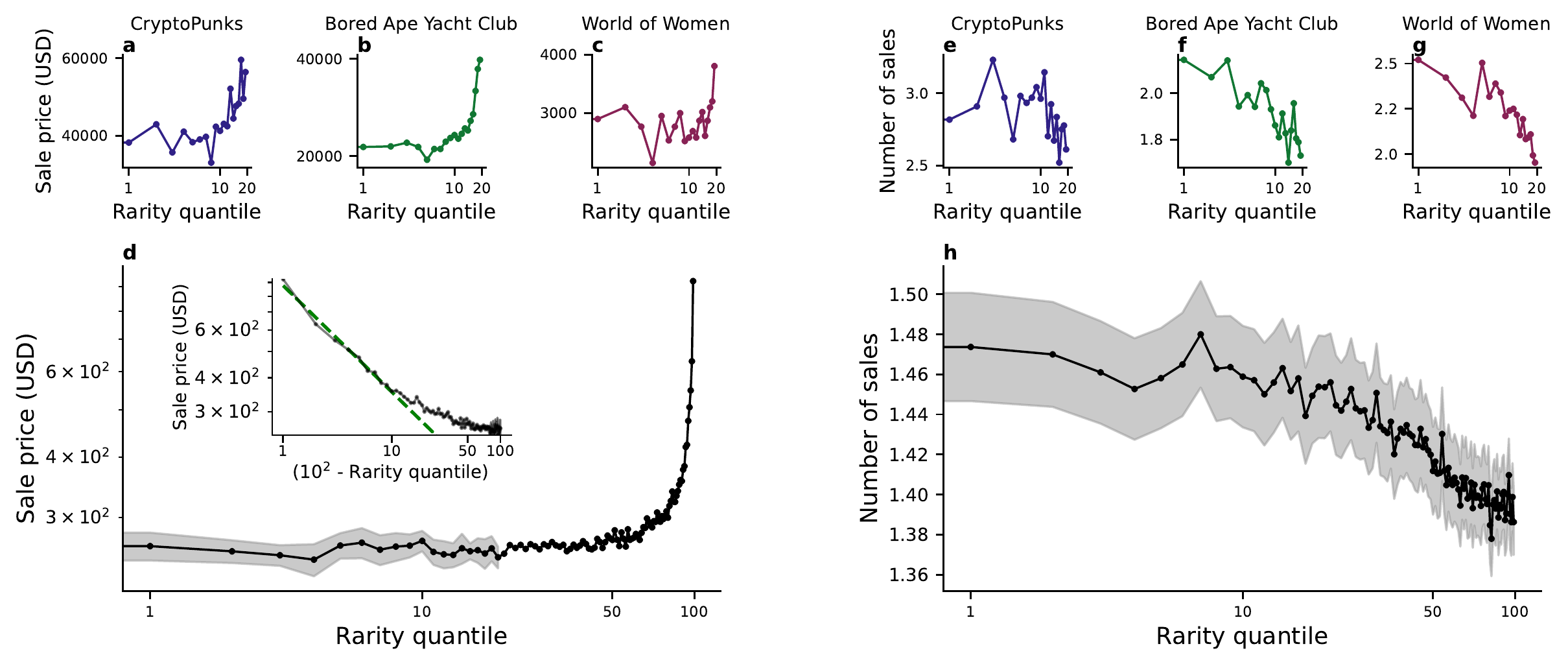}\\
    \caption{\textbf{Rare NFTs have a higher financial value and circulate less on the marketplace - Analysis without the rarest and least rare NFTs.} 
     Median sale price in USD (a-c) and average number of sales (e-g) by rarity quantile (with 20 quantiles considered) for three collections: CryptoPunks (a and e), Bored Ape Yacht Club (b and f), and World Women (c and g). d) Median sale price by rarity quantile (with 100 quantiles considered) considering all collections. Inset: median sale price against the quantity (100-$q$), where $q$ is the rarity quantile, in log-log scale (black line) and the corresponding power law fit (green dashed line). h) Median number of sales by rarity quantile considering all collections. This analysis was performed after discarding the 10\% rarest and least rare NFTs from each collection.}
    \label{fig:impact_coll_tail}
\end{figure}

\begin{figure}[!htb]
    \centering
    \includegraphics[width=0.9\textwidth]{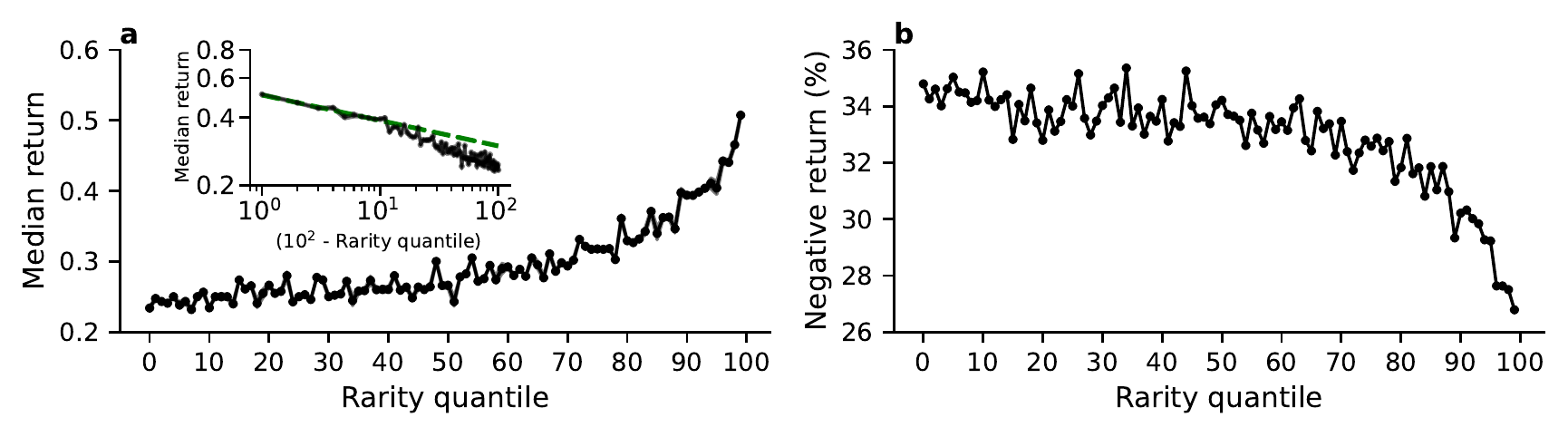}\\
    \caption{\textbf{High rarity leads to higher returns, and a lower chance of a negative return - Analysis without the rarest and least rare NFTs.} a) Median return in USD by rarity quantile. Inset: median return against the quantity (100-$q$), where $q$ is the rarity quantile in log-log scale (black line) and the corresponding power law fit (green dashed line). b) Fraction of sales with negative return in USD by rarity quantile. This analysis was performed after discarding the 10\% rarest and least rare NFTs from each collection.
    }
    \label{fig:amplification_neg_return_tail}
\end{figure}

\end{document}